\documentclass[tighten, preprint]{aastex62}
\pdfoutput=1 
\usepackage{amsmath,amstext}
\usepackage[T1]{fontenc}
\usepackage{txfonts} 
\usepackage[figure,figure*]{hypcap} 
\usepackage{bm}
\usepackage{chngcntr}

\newcommand{\hunmyr}{$100\,\mathrm{Myr}$}

\shorttitle{}
\shortauthors{the IQ (Isolated \& Quiescent) - Collaboratory}

\begin{document}

\title{IQ-Collaboratory 1.1: the Star-Forming Sequence of Simulated Central Galaxies}
\author{ChangHoon Hahn} 
\altaffiliation{hahn.changhoon@gmail.com}
\affil{Lawrence Berkeley National Laboratory, 1 Cyclotron Rd, Berkeley CA 94720, USA}
\affil{Berkeley Center for Cosmological Physics, University of California, Berkeley, CA 94720, USA}
\author{Tjitske K. Starkenburg}  
\affil{Flatiron Institute, 162 Fifth Avenue, New York NY 10010, USA}
\author{Ena Choi}
\affil{Department of Astronomy, Columbia University, 550 West 120th Street, New York, NY 10027, USA}
\author{Romeel Dav{\'e}}
\affil{Institute for Astronomy, Royal Observatory, Edinburgh EH9 3HJ, UK}
\author{Claire~M. Dickey}
\affil{Department of Astronomy, Yale University, New Haven, CT 06520, USA} 
\author{Marla C. Geha}
\affil{Department of Astronomy, Yale University, New Haven, CT 06520, USA} 
\author{Shy Genel} 
\affil{Flatiron Institute, 162 Fifth Avenue, New York NY 10010, USA}
\affil{Columbia Astrophysics Laboratory, Columbia University, 550 West 120th Street, New York, NY 10027, USA}
\author{Christopher C. Hayward} 
\affil{Flatiron Institute, 162 Fifth Avenue, New York NY 10010, USA}
\author{Ariyeh H. Maller}
\affil{Department of Physics, New York City College of Technology, CUNY, 300 Jay St., Brooklyn, NY 11201, USA}
\affil{Department of Astrophysics, American Museum of Natural History, New York, NY 10024, USA}
\author{Nityasri Mandyam}
\affil{Center for Cosmology and Particle Physics, Department of Physics, New York University, New York, NY 10003, USA}
\author{Viraj Pandya} 
\affil{UCO/Lick Observatory, Department of Astronomy and Astrophysics, University of California, Santa Cruz, CA 95064, USA}
\author{Gerg\"{o} Popping} 
\affil{Max-Planck-Institut f\"ur Astronomie, K\"onigstuhl 17, D-69117 Heidelberg, Germany}
\author{Mika Rafieferantsoa} 
\affil{University of the Western Cape, Bellville, Cape Town 7535, South Africa} 
\affil{South African Astronomical Observatory, Observatory, Cape Town 7925, South Africa} 
\affil{Max-Planck-Instit\"ut f\"ur  Astrophysik, D-85748 Garching, Germany}
\author{Rachel S. Somerville}
\affil{Flatiron Institute, 162 Fifth Avenue, New York NY 10010, USA}
\affil{Department of Physics and Astronomy, Rutgers, The State University of New Jersey, 136 Frelinghuysen Rd, Piscataway, NJ 08854, USA}
\author{Jeremy L. Tinker}
\affil{Center for Cosmology and Particle Physics, Department of Physics, New York University, New York, NY 10003, USA}

\begin{abstract}
A tightly correlated star formation rate--stellar mass relation of star 
forming galaxies, or star-forming sequence (SFS), is a key feature in galaxy 
property-space that is predicted by modern galaxy formation models.
We present a flexible data-driven approach for identifying this SFS over
a wide range of star formation rates and stellar masses using Gaussian mixture modeling (GMM). 
Using this method, we present a consistent comparison of the $z{=}0$ SFSs of 
central galaxies in the Illustris, EAGLE, and {\sc Mufasa} hydrodynamic simulations and 
the Santa Cruz semi-analytic model (SC-SAM), alongside data from the Sloan 
Digital Sky Survey. We find, surprisingly, that the amplitude of the SFS 
varies by up to ${\sim}0.7\,\mathrm{dex}$ (factor of ${\sim}5$) among the 
simulations with power-law 
slopes range from 0.7 to 1.2. In addition to the SFS, our GMM method also 
identifies sub-components in the star formation rate--stellar mass relation 
corresponding to star-burst, transitioning, and quiescent sub-populations.  
The hydrodynamic simulations are similarly dominated by SFS and quiescent 
sub-populations unlike the SC-SAM, which predicts substantial fractions of 
transitioning and star-burst galaxies at stellar masses above and below 
$10^{10} M_\sun$, respectively. All of the simulations also produce 
an abundance of low-mass quiescent central galaxies in apparent 
tension with observations. These results illustrate that, even among 
models that well reproduce many observables of the galaxy population, 
the $z{=}0$ SFS and other sub-populations still show marked differences 
that can provide strong constraints on galaxy formation models.
\end{abstract}

\section{Introduction}
\begin{figure}
\begin{center}
\includegraphics[width=0.45\textwidth]{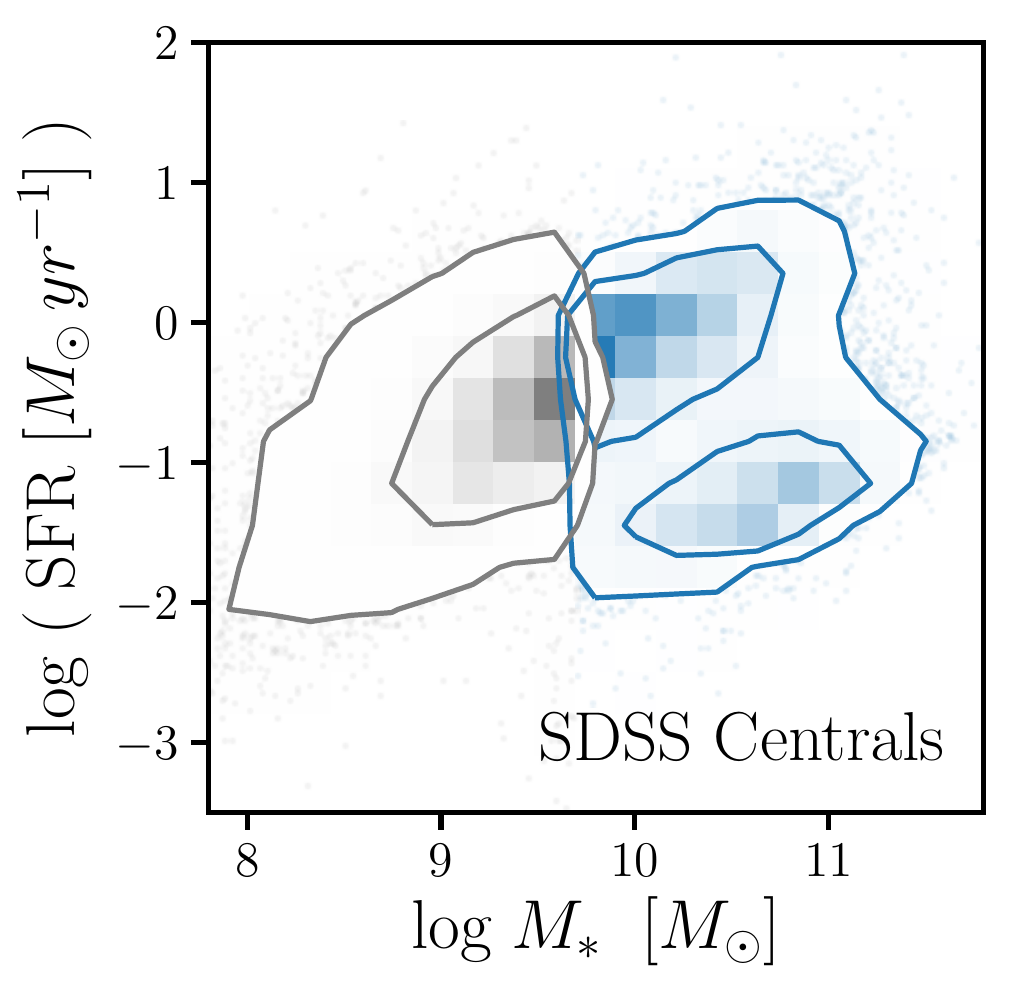} 
\caption{\emph{Star-forming central galaxies in the SDSS 
have a well-defined relationship between their SFRs and stellar masses, 
placing them on the ``star-forming sequence''.} Our SDSS central galaxy 
sample is derived from a volume-limited sample from~\cite{tinker2011} 
at $M_* > 10^{9.7} M_\sun$ (blue) and a low luminosity sample 
from~\cite{geha2012} at $M_* < 10^{9.7} M_\sun$ (gray) described in 
Section~\ref{sec:obvs}.
}
\label{fig:sfrmstar_sdss}
\end{center}
\end{figure}

Large galaxy surveys of the past decade such as the Sloan Digital Sky 
Survey~\citep[SDSS;][]{york2000}, have firmly established the major 
trends of galaxies in the local universe. Galaxies 
broadly fall into two populations: quiescent galaxies with little star
formation that are red in color with elliptical morphologies and star 
forming galaxies with significant star formation that are blue in color 
with disk-like morphologies 
(\citealt{kauffmann2003, blanton2003, baldry2006, taylor2009, moustakas2013}; 
see~\citealt{blanton2009} and references therein). 
Star-forming galaxies, furthermore, are found to have a tight relationship 
between their star formation rates (SFR) and stellar masses placing them
on the so-called ``star-forming sequence''~\citep[hereafter SFS; \emph{e.g.}][see also Figure~\ref{fig:sfrmstar_sdss}]{noeske2007, daddi2007, salim2007}.


In fact, this sequence of star-forming galaxies is found in observations 
well beyond the local universe out to $z > 2$~\citep{wang2013, leja2015, schreiber2015}.
But more than its persistence, the SFS plays a crucial role in characterizing 
the evolving galaxy population. Although its importance is 
contested~\citep{kelson2014,abramson2016}, the most dramatic transformations 
of galaxies over the past $10\,\mathrm{Gyr}$ can be described by the SFS. 
For instance, the decline in the number density of massive 
star-forming galaxies and the accompanying growth in number density of 
quiescent galaxies reflects the cessation of star formation in 
star-forming galaxies migrating off of the 
SFS~\citep{blanton2006, borch2006, bundy2006, moustakas2013}. 
Similarly, the cosmic decline in star formation~\citep{hopkins2006,
behroozi2013a, madau2014} reflects the overall decline of star 
formation of the SFS~\citep{schreiber2015}. 

Galaxy formation models \emph{qualitatively} reproduce the SFS and 
similar global relations of galaxy properties at $z\sim0$ 
and provide insights into the key physical processes governing those relations 
(\emph{e.g.}~\citealt{vogelsberger2014,genel2014, schaye2015, somerville2015a, dave2017}; 
for a recent review see~\citealt{somerville2015}). These hydrodynamic and 
semi-analytic simulations each seek to capture the complex physics of 
gas heating and cooling, star formation, stellar feedback, chemical 
evolution, black hole formation and evolution, and feedback from active 
galactic nuclei (AGN) using their distinct sub-grid model prescriptions.
Many works have already compared the simulations considered in this paper to 
observations of, for example, galaxy masses, colors, and star formation 
rates~\citep[\emph{e.g.}][]{vogelsberger2014, genel2014, torrey2014, sparre2015, schaye2015, bluck2016, dave2017, somerville2015}. 
These works, however, primarily focus on comparing one specific simulated 
galaxy sample to one or a few observational datasets. Extending such 
comparisons to include multiple simulations, observations, and a 
consistent framework for comparing the data-sets would allow us to make 
detailed comparison of the different sub-grid models and thereby provide 
key constraints on the physics that govern galaxy formation and evolution.  

The SFS, given its prominence, naturally presents itself as a key feature 
in the data-space of galaxy properties to compare across both observations 
and simulations. Moreover, with the important role it plays in characterizing 
the evolving galaxy population, the SFS provides a way to interpret and 
understand the different galaxy subpopulations and the processes that create them.
Two main challenges lie in conducting a meaningful comparison of the SFS. First 
is the lack of a flexible and data-driven method for identifying the SFS across 
different datasets. In fact, inconsistencies in how the SFS is identified 
have incorrectly led to agreement among simulations and 
observations~\citep[\emph{e.g.}][see Appendix~\ref{app:literature}]{somerville2015}. 
The other challenge is the difference in methodology for 
deriving galaxy properties (such as SFR, $M_*$), which even for the same 
data-set dramatically impacts the SFS~\citep[\emph{e.g.}][]{speagle2014}. 
In this paper we address the first challenge by presenting a flexible, 
data-driven method for identifying the SFS. We then use this method to compare 
the $z = 0$ central galaxy populations of the Illustris, EAGLE, and {\sc Mufasa} 
hydrodynamic simulations and the Santa Cruz Semi-Analytic Model (SC-SAM), alongside 
observations from SDSS. 

In Section~\ref{sec:galsims}, we describe the data-sets from simulations
and observations and how we specifically select our galaxy sample. 
Then in Section~\ref{sec:sfmsfit}, we describe how we identify the SFS 
with a data-driven approach using Gaussian mixture modeling. We present 
the resulting SFSs from the simulations and observations in 
Section~\ref{sec:results} and compare the galaxy populations
of the simulations and observations. Finally, we conclude and summarize
the results of our comparison in Section~\ref{sec:summary}.
This paper is the first in a series, initialized by the IQ (Isolated 
\& Quiescent)-Collaboratory, which aims to improve our understanding 
of quenching processes by comparing isolated star-forming 
and quiescent galaxies in simulations and observations. This first 
project in the IQ-Collaboratory focuses on the star-forming and quiescent 
galaxy populations at $z\sim 0$. Additional projects will focus on galaxy 
populations at the peak of cosmic star formation (Choi et al. in prep.), 
and the gas content of star-forming and quenched galaxies (Emerick et al. 
in prep.).  
In the subsequent paper of this project (IQ 1.2), we will address the challenges 
in measured galaxy properties by forward modeling mock spectra of simulated 
galaxies and measuring their properties in the same manner as observations 
(Starkenburg et al. in prep).

\section{The Galaxy Samples} \label{sec:galsims}
In this work, our main focus is to compare simulated central galaxies 
from four large-scale cosmological simulations: three hydrodynamic 
(Illustris, EAGLE, and {\sc Mufasa}) and one semi-analytic (SC-SAM). 
A consistent comparison requires consistently defined galaxy properties
across the simulations. For all of the simulated galaxies we derive 
their stellar masses using the same 
definition and their SFRs on two timescales: instantaneous and averaged over 
$100\,\mathrm{Myr}$. SFR on these timescales correspond to $H{\alpha}$ 
and $UV$ based SFR measurements, which represent the formation of young 
stars with ages ${\lesssim}10\,\mathrm{Myr}$ and star formation in the 
last ${\sim}100\,\mathrm{Myr}$~\citep[e.g.][]{kennicutt2012}, 
respectively. We use instantaneous SFR, instead of SFR averaged over 
$10\,\mathrm{Myr}$, to minimize resolution effects in hydrodynamic simulations 
on such short timescales (Appendix~\ref{app:zerosfr}).

In the hydrodynamic simulations, we derive the instantaneous SFRs from 
the rate of star formation in the dense and cold gas and the $100\,\mathrm{Myr}$ 
averaged SFRs from the ages, or formation times, of star particles 
in the galaxies. For the semi-analytic model, we derive the 
instantaneous SFR using the Kennicutt-Schmidt relation for molecular 
hydrogen~\citep[based on][]{bigiel2008} and the derived H$_2$ surface 
density in radial bins. We derive the $100\,\mathrm{Myr}$ averaged SFRs 
from the total stellar mass formed in the galaxies, which is outputted 
from the model every $10\,\mathrm{Myr}$. 
Meanwhile, for the stellar mass of the simulated galaxies, we 
use the total stellar mass within the host halos, discounting the 
stellar mass in any subhalo within the halo. While stellar masses 
defined within some effective radius is better suited for comparison 
to observations, we use the total stellar mass within the halo because
different sub-grid models can significantly alter galaxy sizes.
Furthermore, although halos are identified differently in the simulations, 
nearly all the stellar mass is in the center of halos. Therefore, the 
stellar masses are consistently defined among the simulations. 

From the SFRs and stellar masses, we derive the specific-SFRs of the galaxies 
as $\log\,\mathrm{SSFR} = \log\,\mathrm{SFR} - \log\,M_*$. Due to the numerical 
and resolution effects a significant number of galaxies in the hydrodynamic 
simulations have $100\,\mathrm{Myr}$ averaged ``SFR$=0$'', when their SFRs 
are below the resolution limit of the simulations. We consider these galaxies 
to have ``unmeasurably low SFRs''. For the instantaneous SFRs and both SFRs for 
the SC-SAM, we analogously consider $\log\,\mathrm{SFR} < -4\ M_{\sun} \mathrm{yr}^{-1}$ 
as "unmeasureably low SFR". We discuss in Appendix~\ref{app:zerosfr}, how 
we treat the effect of spatial, mass, and temporal resolution of the simulations, 
which can impact SFR, in further detail.

In the rest of this section we provide a brief description of the 
Illustris, EAGLE, {\sc Mufasa}, and SC-SAM simulations and each 
of their key sub-grid and feedback prescriptions. In addition, we briefly 
describe the
SDSS galaxy sample, which we include for reference in Section~\ref{sec:obvs}.
Lastly, we describe how we consistently identify central galaxies among 
the simulations and observations in Section~\ref{sec:central}.

\subsection{Illustris}
The Illustris 
simulation\footnote{\url{http://www.illustris-project.org}}~(\citealt{vogelsberger2014,genel2014}; public data release~\citealt{nelson2015a}) 
evolves a cosmological volume of $(106\ \rm{Mpc})^3$ with a uniform 
baryonic mass resolution of $1.26\times10^6M_{\sun}$ using the {\sc Arepo} 
moving-mesh code~\citep{springel2010}. 
It employs sub-grid models for star-formation~\citep{springel2003},
Bondi-like supermassive black hole (SMBH) accretion, a phenomenological 
model for galactic winds~\citep{oppenheimer2006}, and two main modes for energy injection 
from SMBHs~\citep[see][]{vogelsberger2013}. When gas accretion onto the 
SMBH occurs at Eddington ratios $>0.05$, thermal energy is injected 
continuously in  the local environment of the SMBH. At lower accretion 
rates, the  energy injection occurs in bursts at large distances from the 
SMBH, generating hot bubbles in the intracluster medium~\citep{sijacki2007}.
Previous works discussing aspects of the SFS and/or quenching in Illustris 
include \citet{genel2014, vogelsberger2014, sparre2015, bluck2016, terrazas2017}. 

\subsection{EAGLE}
The Virgo Consortium's Evolution and Assembly of GaLaxies and their 
Environment (EAGLE) project\footnote{\url{http://www.eaglesim.org}}~\citep{schaye2015, crain2015} is a publicly available~\citep{mcalpine2016} suite 
of cosmological, hydrodynamic simulations of a standard 
$\Lambda$ cold dark matter universe. Of the simulations, we use 
L0100Ref, which has a volume of $(100\,\mathrm{comoving\,Mpc})^3$ and 
baryonic mass resolution of $1.81\times 10^6M_{\sun}$. 
It uses {\sc Anarchy} (Dalla Vecchia et al. in prep.; 
see also Appendix A of \citealt{schaye2015} and \citealt{schaller2015}), 
which is a modified version of the ${\sc Gadget}$ 3 $N$-body/SPH code~\citep{springel2005} 
that includes modifications to the SPH formulation, time stepping, and 
sub-grid physics. The sub-grid model for feedback from massive stars 
and AGN is based on thermal energy injection in the ISM~\citep{dallavecchia2012}.  
The simulations resolve galaxies above $M_* > 10^{8} M_\sun$. 
The SFR--$M_*$ relation and quiescent fractions in EAGLE have been 
previously discussed in~\citet{furlong2015, trayford2015, trayford2017}. 

\subsection{{\sc Mufasa}} \label{sec:mufasa}
{\sc Mufasa} is a hydrodynamic simulation with a box size of 
$(50\,h^{-1}\ {\rm Mpc})^3$ and particle masses of $9.6 \times 10^7\ M_{\sun}$ 
and $1.82 \times 10^7\ M_{\sun}$ for dark matter and baryons, respectively. 
It  uses {\sc Gizmo}, a code built on {\sc Gadget} that uses the Meshless 
Finite Mass hydrodynamics method~\citep{hopkins2015a} rather than SPH.  
{\sc Mufasa} includes star formation via a Kennicutt-Schmidt law based on the  molecular hydrogen 
density as computed using the sub-grid recipe in~\cite{krumholz2011}. 
It also includes two-phase kinetic outflows with scalings as predicted 
in the Feedback in Realistic Environments (FIRE) simulations~\citep{muratov2015}.
Finally, it quenches massive galaxies by keeping all non-self shielded 
gas within halos above a mass of $M_q>(1+0.48 z)\,10^{12}M_\sun$ 
near the halos' virial temperature~\citep{gabor2015, mitra2015}. The stellar mass 
function, gas and metal content of galaxies, and color-mass 
diagram of {\sc Mufasa} have been previously discussed 
in~\citet{dave2016,dave2017,dave2017b}.

\subsection{Santa Cruz Semi-Analytic Model} \label{sec:scsam}
The `Santa Cruz' SAM (SC-SAM) is a semi-analytic model run on 
merger trees from a ($100$ comoving Mpc/$h$)$^3$ subvolume of the 
Bolshoi--Planck dark matter only $N$-body simulations~\citep{rodriguez-puebla2016}. 
The Bolshoi--Planck simulations have particle masses of $1.5 \times 10^8 M_\sun$. 
The model includes schematic prescriptions for gas heating and 
cooling, multi-phase gas partitioning, star formation, chemical 
evolution, feedback from stars, supernovae and SMBHs, the sizes 
of galactic disks and 
bulges, and merger-induced starbursts and structural transformations. 
The SC-SAM was first presented in~\cite{somerville1999} and 
\cite{somerville2001}, with significant updates described 
in~\cite{somerville2008, somerville2008a, somerville2012, porter2014, 
popping2014, somerville2015a}. In this work, we use the version of 
the SC-SAM described in~\cite{popping2014} and~\cite{somerville2015a}, 
which includes the \cite{gnedin2011} recipe for partitioning multi-phase
gas into HI, H$_2$ and HII 
based on the dark matter resolution limit,
we focus our analysis on halos with $M_h{>}10^{11} M_\sun$.
Since this roughly corresponds to $M_*{\sim}10^{8.5} M_\sun$ at 
$z\sim 0$, we impose a conservative cut of $M_* > 10^{8.8}M_\sun$. 
The properties of the SC-SAM galaxy population, 
such as the quiescent fraction have been previously discussed 
in~\cite{brennan2015,somerville2015a,somerville2015,brennan2017,pandya2017}.


\subsection{Observed SDSS Galaxies} \label{sec:obvs}
As a reference to our comparison of the simulated galaxies,  
we include SDSS galaxies from two samples: a 
$M_*{>}10^{9.7} M_\sun$ Data Release 7~\citep[DR7;][]{abazajian2009} sample 
and a $M_* < 10^{9.7} M_\sun$ Data Release 8~\citep[DR8;][]{aihara2011} sample
(blue and gray in Figure~\ref{fig:sfrmstar_sdss}). 
At high masses, we use the volume-limited galaxy sample 
from~\cite{tinker2011} constructed from the NYU Value-Added Galaxy 
Catalog~\citep[VAGC;][]{blanton2005}. It has $M_r - 5\log(h) < -18$
and is complete for $M_* > 10^{9.7} M_\sun$. For further details, 
we refer readers to~\cite{tinker2011,wetzel2013,hahn2017}. 

At lower stellar masses, we use the isolated dwarf galaxy sample 
of \citet{geha2012} from the NASA Sloan Atlas (NSA), a reprocessing 
of SDSS DR8. The NSA is optimized for low-luminosity objects and 
relies on the improved background subtraction technique of 
\cite{blanton2011}. The catalog extends to $z \approx 0.055$ and 
includes re-calibrated spectroscopy~\citep{yan2011,yan2012} 
with much smaller errors\footnote{This recalibration, however, is mostly 
relevant only at small equivalent width values and hence does not 
largely affect galaxies on the SFS.}.

For both SDSS subsamples, the stellar masses are estimated using the 
\citet{blanton2007} $\mathtt{kcorrect}$ code, which assumes a 
\cite{chabrier2003} IMF. The SFRs are from the current release 
of~\citet{brinchmann2004}\footnote{http://www.mpa-garching.mpg.de/SDSS/DR7/}, 
where they are derived using the~\cite{bruzuala.1993} model with the 
\cite{charlot2000} dust prescription and CLOUDY \citep[version C90.04;][]{ferland1996}
emission line modeling. For galaxies classified as having an AGN or a 
composite spectrum, the SFR is measured from the $D_n4000$ index~\citep{balogh1998}. 
Additionally, for star-forming galaxies that have low S/N spectra, the SFR 
is derived from the $H{\alpha}$ luminosity~\citep{brinchmann2004}. 
We emphasize that SSFRs $\lesssim 10^{-12} \mathrm{yr}^{-1}$ should only be 
considered upper limits to the true value~\citep{salim2007}.
Given the disparate methods used for the SFR measurements, the SFRs in 
the SDSS sample do not entirely correspond to either the instantaneous 
or $100\,\mathrm{Myr}$ averaged SFRs of the simulations. Consequently, 
in this work we compare the simulations on both timescales and refrain 
from detailed comparisons to SDSS. 


\subsection{Identifying Central Galaxies} \label{sec:central}
Measurements of the quiescent fraction~\citep[\emph{e.g.}][]{baldry2006,peng2010,hahn2015}
and star formation quenching timescale~\citep{wetzel2013,hahn2017} 
suggest, whether a galaxy is a satellite or central galaxy influences 
its star formation rate. There may also be significant differences 
between the SFSs of central versus satellite galaxies~\citep{wang2018}. 
In this paper \emph{we focus solely on the central galaxies, which 
constitute the majority of massive galaxies ($M_* > 10^{9.5}M_\sun$) at $z \sim 0$}. 

Central classification, despite its importance, is often 
heterogeneously defined in the literature. Among simulations, the 
classification depends on the definition of halo properties, 
and thus on the underlying halo finders. EAGLE and Illustris use 
$\mathtt{SUBFIND}$~\citep{springel2001}, where halos are defined as 
locally overdense, gravitationally bound (sub)structures within a 
connected region selected through a 
friend-of-friends~\citep[FOF;][]{davis1985} group finder. {\sc Mufasa} 
and SC-SAM, meanwhile, use 
$\mathtt{ROCKSTAR}$~\citep{behroozi2013}, which defines halos using a
hierarchical phase-space based FOF technique and seeks to maximize 
the consistency of the halo through time. In addition, these central 
classifications also use information of the underlying dark 
matter --- information {\em not} available in observations. Therefore, 
we identify central galaxies in all simulations consistently using 
the~\cite{tinker2011} group finder, designed to identify satellite/centrals 
in observations. 

The~\cite{tinker2011} group finder is a halo-based algorithm that uses 
the abundance matching ansatz to iteratively assign halo masses to groups. 
It assigns a tentative halo mass to each galaxy by matching the abundance 
of the objects. Then starting with the most massive galaxy, nearby lower
mass galaxies are assigned a probability of being a satellite. Once all 
the galaxies are assigned to a group, the halo masses of the central galaxies 
are updated by abundance matching with the total stellar mass in the groups. 
This entire process is repeated until convergence. In the resulting catalog, 
every group contains one central galaxy, which by definition is the 
most massive, and a group can contain zero, one, or many satellites.
For a detailed description we refer readers to~\cite{tinker2011,wetzel2012}. 

Overall, we find good agreement between the central classifications of 
the group finder with respect to that of the simulations 
 with purities of
$99\%, 93\%, 84\%$, and $97\%$ and completenesses of $86\%$, $89\%$, 
$91\%$, and $85\%$ for the Illustris, EAGLE, {\sc Mufasa} and SC-SAM 
simulations respectively. Differences in the purity and completeness for the 
simulations is likely due to the different halo finders used in the 
simulations. 
We find no significant stellar mass dependence in the purities. 
As expected from the high purity and completeness, when we perform our 
analysis using the centrals and identified by the dark matter halos, we 
find no significant differences. 
In the next section, we proceed to fitting the star-forming sequence 
of simulated \emph{central} galaxies. 

\begin{figure*}
\begin{center}
\includegraphics[width=0.95\textwidth]{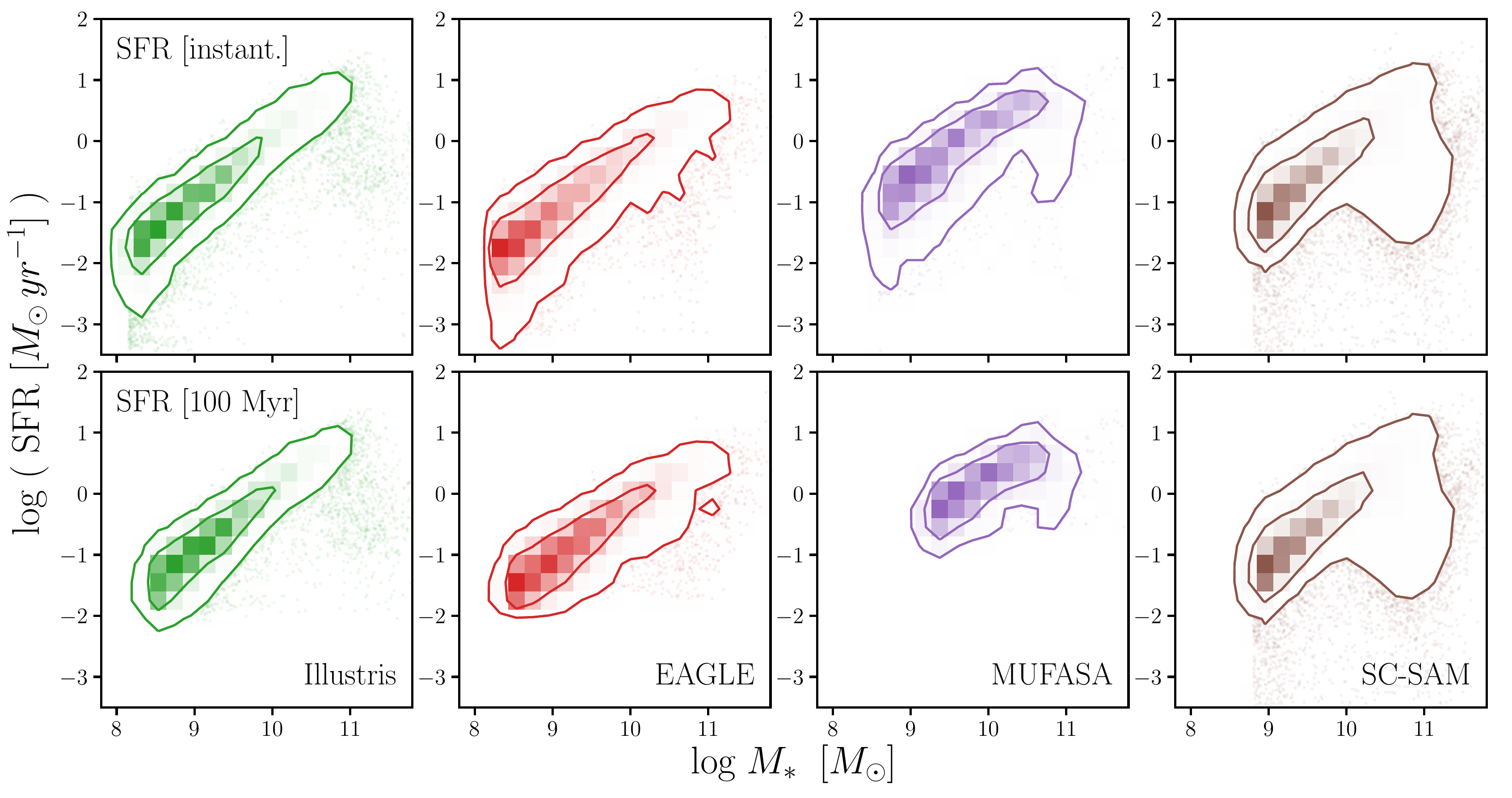} 
\caption{The SFR--$M_*$ relations of central galaxies from 
the Illustris (green), EAGLE (red), and {\sc Mufasa} (purple) hydrodynamic 
simulations and the SC-SAM (brown) at $z=0$. The top panels use instantaneous 
SFRs while the bottom panels use SFRs averaged over $100\,\mathrm{Myr}$. The 
contours in each panel mark the $68\%$ and $95\%$ confidence intervals of the 
SFR--$M_*$ distribution. We describe the simulations and how we derive consistent 
SFRs and stellar masses in Section~\ref{sec:galsims}. \emph{The SFR--$M_*$ relations 
reveal star-forming sequences in all of the simulations.}} 
\label{fig:sfrmstar}
\end{center}
\end{figure*}
\begin{figure*}
\begin{center}
\includegraphics[width = 0.9\textwidth]{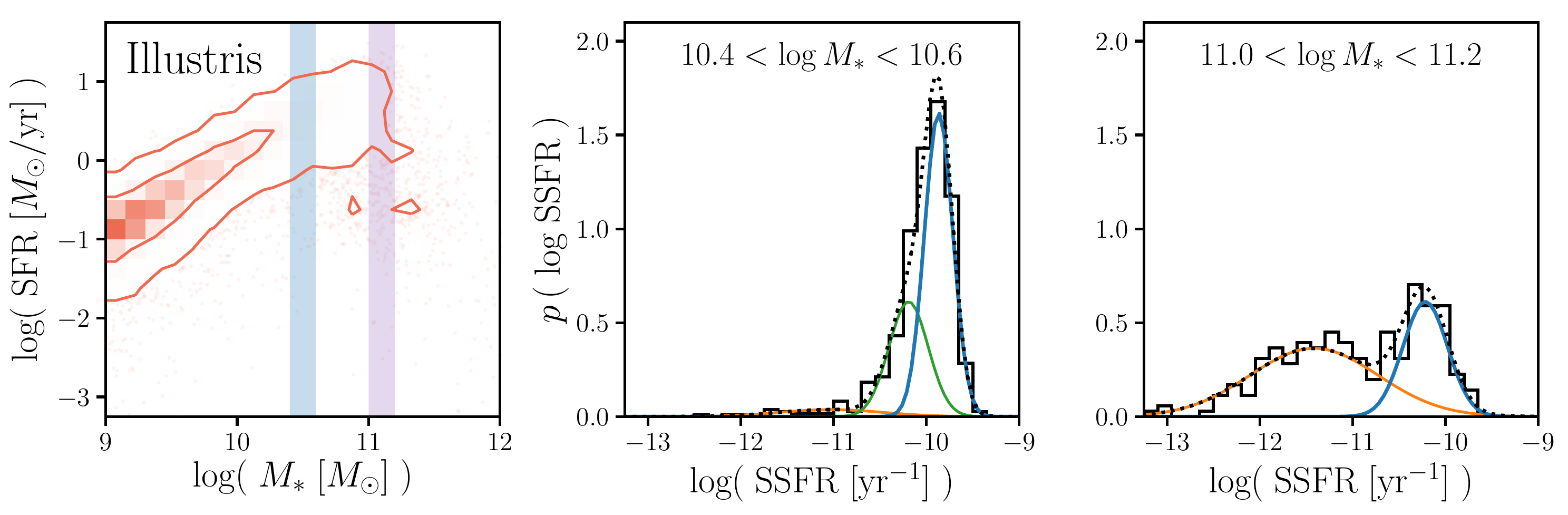} 
\caption{
We illustrate our 
{GMM based method for identifying the SFS of} Illustris central galaxies in two 
stellar mass bins highlighted on the SFR--$M_*$ relation of the left panel: 
$10.4 < \log\,M_* < 10.6$ and $11.0 < \log\,M_* < 11.2$. We compare the SSFR 
distributions, $p(\log\,\mathrm{SSFR})$, in the two stellar mass bins to their 
best-fit GMMs (right panels). The $p(\log\,\mathrm{SSFR})$ in the center panel is best described by a 
GMM with three components (orange, green, and blue) while the
$p(\log\,\mathrm{SSFR})$ in the right panel is best described by 
a GMM with two components (orange and blue). The SFS components of the 
best-fit GMMs are plotted in blue. \emph{Our GMM method provides
a flexible and data-driven method of identifying the SFS in a wide variety 
of SSFR distributions without hard assumptions or cuts to the sample.}
}\label{fig:fitdemo}
\end{center}
\end{figure*}

\section{Identifying the Star-Forming Sequence}\label{sec:sfmsfit}
We present the SFR--$M_*$ relation of central galaxies from the 
observations and simulations of Section~\ref{sec:galsims} in 
Figures~\ref{fig:sfrmstar_sdss} and~\ref{fig:sfrmstar}. For both 
instantaneous and $100\,\mathrm{Myr}$ SFRs (top/bottom),
in both simulations and observations, and over four orders of magnitude 
in SFR and stellar mass, \emph{the SFR and $M_*$ of star-forming galaxies 
lie on a well-defined SFS.} Despite its universality, in detail, the different 
datasets give rise to different SFR--$M_*$ distributions, which makes the 
SFS difficult to consistently and meaningfully quantify. 
So far in the literature, a wide variety of fitting methods has been applied to 
data --- even in a single comparison (see Appendix~\ref{app:literature}). 
For example, in \cite{lee2015} and some of the fits in~\cite{somerville2015}
the SFS is fit using median $\log\,\mathrm{SFR}$s of galaxies after some 
color-color or SSFR cut to the sample. Other SFSs in~\cite{somerville2015} 
are fit using the median $\log\,\mathrm{SFR}$s of the entire sample. 
\cite{bluck2016} fit the SFS using median $\log\mathrm{SFR}$s of low mass 
galaxies ($M_* < 10^{10}M_\sun$) and extrapolate to higher masses. 
Other recent works in the literature have opted for more sophisticated 
methods such as fitting a three-component Gaussian~\citep{bisigello2018} 
or a zero-inflated negative binomial distribution~\citep{feldmann2017}. 

All of these methods require arbitrary assumptions or hard cuts to the 
sample. More importantly, for such methods, different assumptions 
or cuts produce different SFSs and inconsistent assumptions and cuts 
can result in misleading SFS comparisons (Appendix~\ref{app:literature}).
Identifying the SFS also requires flexibility in accounting 
for the different features in the galaxy property space over a wide SFR or 
$M_*$ range and in different simulations and observations.
In an effort to better fit the SFS from a wide variety of SFR--$M_*$ 
distributions and to relax the assumptions and cuts imposed on the data, 
\emph{we present a flexible and data-driven method for identifying the SFS 
that makes use of Gaussian Mixture Models}.

\subsection{Using Gaussian Mixture Models} \label{sec:gmm}
Gaussian mixture models (hereafter GMM), and mixture models in general, provide 
a probabilistic way of describing the distribution of a population by 
identifying subpopulations from the data~\citep[][]{Press:1992:NRC:148286, 9780471006268}.
Besides their extensive use in machine learning and statistics, 
GMMs have also been used in a wide range of astronomical analyses~\citep[\emph{e.g.}][]{bovy2011,lee2012,taylor2015}. 
Since identifying the subpopulation of star-forming galaxies from the overall
galaxy population is equivalent to identifying the SFS, GMMs provides a 
well-motivated, data-driven, and effective method to tackle the problem. 

A GMM, more precisely, is a weighted sum of $k$ Gaussian component densities 
\begin{equation} \label{eq:gmm}
\hat{p}(x;\bm{\theta}) = \sum\limits_{i=1}^{k} \pi_i \, \mathcal{N}(x; \bm{\theta}_i),
\end{equation}
which can be used to estimate the density. The weights, $\pi_i$, mean, and 
variance $\bm{\theta}_i=\{\mu_i, \sigma_i\}$ of the components are free 
parameters. For a given data set $\{x_1, ..., x_n\}$, these 
parameters are most commonly estimated through the expectation-maximization 
algorithm~\citep[EM;][]{dempster1977,neal1998}. 

Starting with randomly assigned $\bm{\theta}_{i}^0$ to the $k$ GMM components, 
the EM algorithm iterates between two steps. First, for every data point, 
$x_i$, the algorithm computes for a probability of $x_i$ being generated by 
each GMM component. These probabilities act as assignment weights to each of
the components. Next, based on these weights, $\bm{\theta}_i^t$ of the components 
are updated to $\bm{\theta}_i^{t+1}$ to maximize the likelihood of the assigned 
data. $\pi_i$ are also updated by summing up the assignment weights and 
normalizing the sum by the total number of data points. These steps are 
repeated until $p(\{x_1, ..., x_n\} ; \bm{\theta}_t)$ converges. Instead of 
starting with randomly assigning $\bm{\theta}_{i}^0$, we initiate our EM algorithm 
using a $k$-means clustering algorithm~\citep{lloyd1982}, more specifically 
we use the $k$-$\mathtt{means}$++ algorithm~\citep{arthur2007}. 

For actually identifying the SFS, we first divide the galaxy 
sample into stellar mass bins of some width $\Delta \log M_∗$. In this paper 
we use bins of $\Delta \log M_* = 0.2\ \mathrm{dex}$; however, this 
choice does not significantly impact the final SFS. For each stellar 
mass bin, if there are more than $N_\mathrm{thresh}{=}100$ galaxies in the bin, 
we fit the SSFR distribution using GMMs with $k{=}1$ to 3 components with 
parameters determined from the EM algorithm described above. 
For the SDSS galaxy sample and the hydrodynamic simulations, even when we 
allow for more than 3 components, the best-fit GMMs have $k\leq3$. Hence, the 
choice of $k\leq3$ does not significantly impact the results of this work.
Out of the three ($k\leq3$) GMMs, we select the one with the lowest Bayesian 
Information Criteria~\citep[BIC;][]{schwarz1978} as our ``best-fit'' model. 
BIC is often used in conjunction with GMMs~\citep[\emph{e.g.}][]{leroux1992,roeder1997,fraley1998,steele2010performance} 
and also more generally for model selection in 
astronomy~\citep[\emph{e.g.}][]{liddle2007,broderick2011,vakili2016}.
In addition to the likelihood, BIC introduces a penalty term for the number
of parameters in the model. This way, using BIC not only finds a good fit to 
the data, but it also addresses the concern of over-fitting. 

Given the best-fit GMM, we next identify the SFS components in each $\log M_*$
bin. We start from the lowest $\log M_*$ bin, where we take the component with 
the largest weight as the SFS component. Then in the next higher $\log M_*$ bin
we identify the component with the largest weight. If this 
component has a mean within $0.5\,\mathrm{dex}$ of the previous lower $\log M_*$ 
bin SFS component mean, we identify this component as the SFS. Otherwise, we 
discard it and determine whether the component with the next highest weight 
is within $0.5\,\mathrm{dex}$ of the previous SFS component mean. We 
repeat this until we either identify a SFS component or, if no 
component is within $0.5\,\mathrm{dex}$ of the previous SFS component mean,
conclude that no SFS component is in the $\log M_*$ bin. We repeat this 
procedure recursively for all the $\log M_*$ bins. This scheme takes advantage 
of the bimodality in the SSFR distributions and assumes that the SFS forms a 
relatively continuous sequence. In Appendix~\ref{app:gmm_pssfr}, we present a 
detailed comparison of the GMM fits to the SSFR distributions of the simulations 
and discuss the advantages of our method in further detail.



In Figure~\ref{fig:fitdemo}, we illustrate our GMM based method for identifying the SFS of the Illustris central galaxies 
in two stellar mass ranges highlighted in the left panel: $10.4 < \log\,M_* < 10.6$ (center) 
and $11.0 < \log\,M_* < 11.2$ (right). For the two stellar mass bins, 
we compare the SSFR distributions of the bins to the components of the 
best-fit GMMs derived from our method. The SFS components of the best-fit 
GMMs are plotted in blue. The SSFR distribution of the center panel is best 
described by a GMM with three components while the SSFR distribution 
in the right  panel is best described by a GMM with only two components.
These comparisons highlights the flexibility and effectiveness of our 
method in identifying the SFS for different SSFR  distributions. 
Our code for identifying the SFS makes use of the following software: 
{\em astroML}~\citep{astroML}, {\em astropy}~\citep{astropy:2013,theastropycollaboration2018}, 
{\em matplotlib}~\citep{Hunter:2007}, {\em numpy}~\citep{numpy:2011}, 
{\em scipy}~\citep{scipy:2001}, and {\em scikit-learn}~\citep{scikit-learn:2011}. 
All of the code is publicly available at \url{https://github.com/changhoonhahn/LetsTalkAboutQuench}.

\begin{figure*}
\begin{center}
\includegraphics[width = 0.8\textwidth]{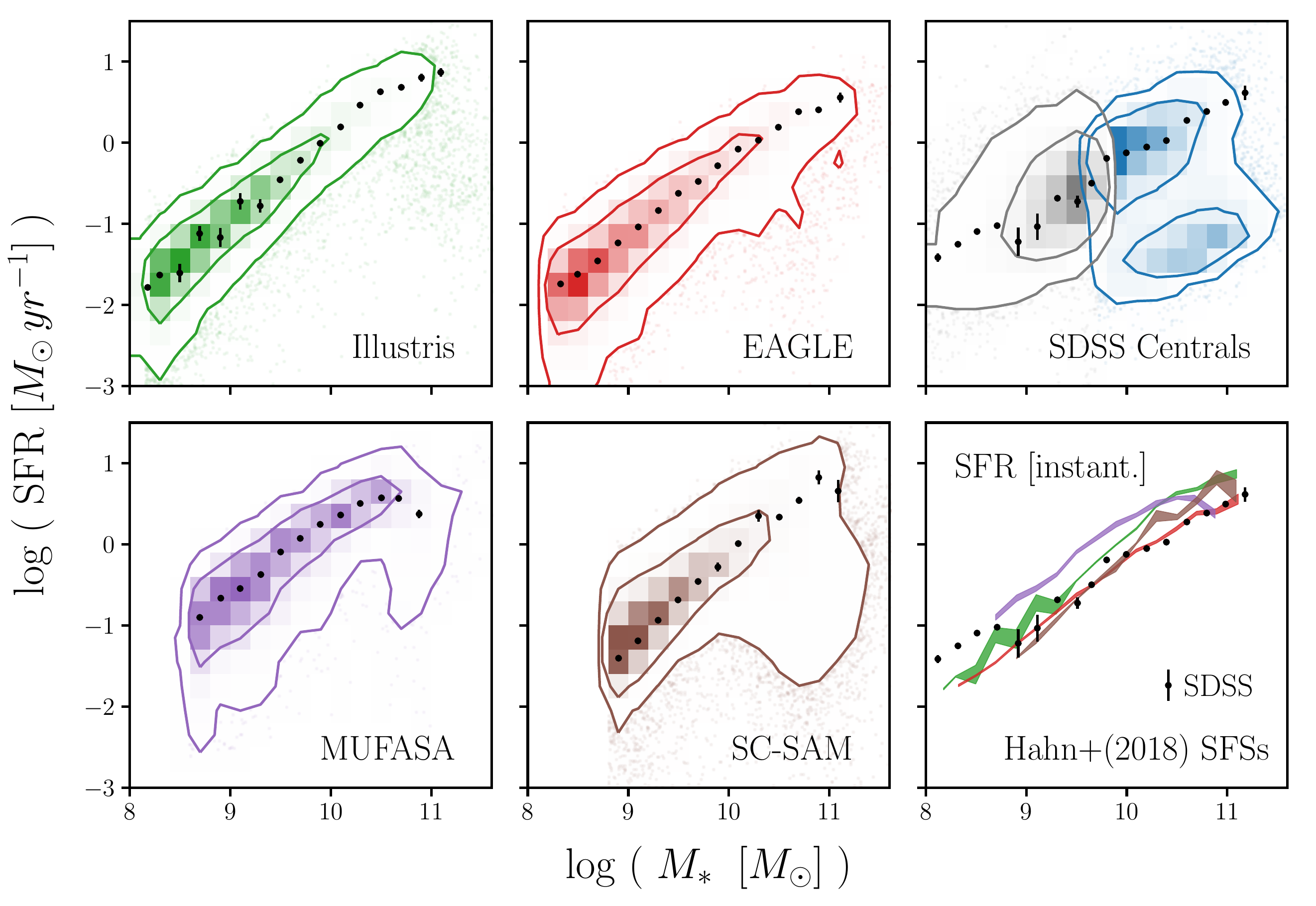} 
\caption{The SFSs of the central galaxies in the Illustris, EAGLE, {\sc Mufasa}, 
    and SC-SAM simulations as identified by our GMM based method (Section~\ref{sec:sfmsfit}).
    The SFSs above are identified from the instantaneous SFR--$M_*$ relation. 
    The uncertainties of the SFSs are derived using bootstrap resampling 
    and marked by the error bars. For reference, we include the SFS of the SDSS 
    sample in the top right panel and the bottom right panel (black). 
    When we compare the \emph{SFSs of the simulations we find that they have significantly different slopes and their amplitudes vary by up to ${\sim}0.7\,\mathrm{dex}$, 
    factor of ${\sim}5$} 
    (bottom right).} \label{fig:sfmsfit_inst}
\end{center}
\end{figure*}

\begin{figure*}
\begin{center}
\includegraphics[width = 0.8\textwidth]{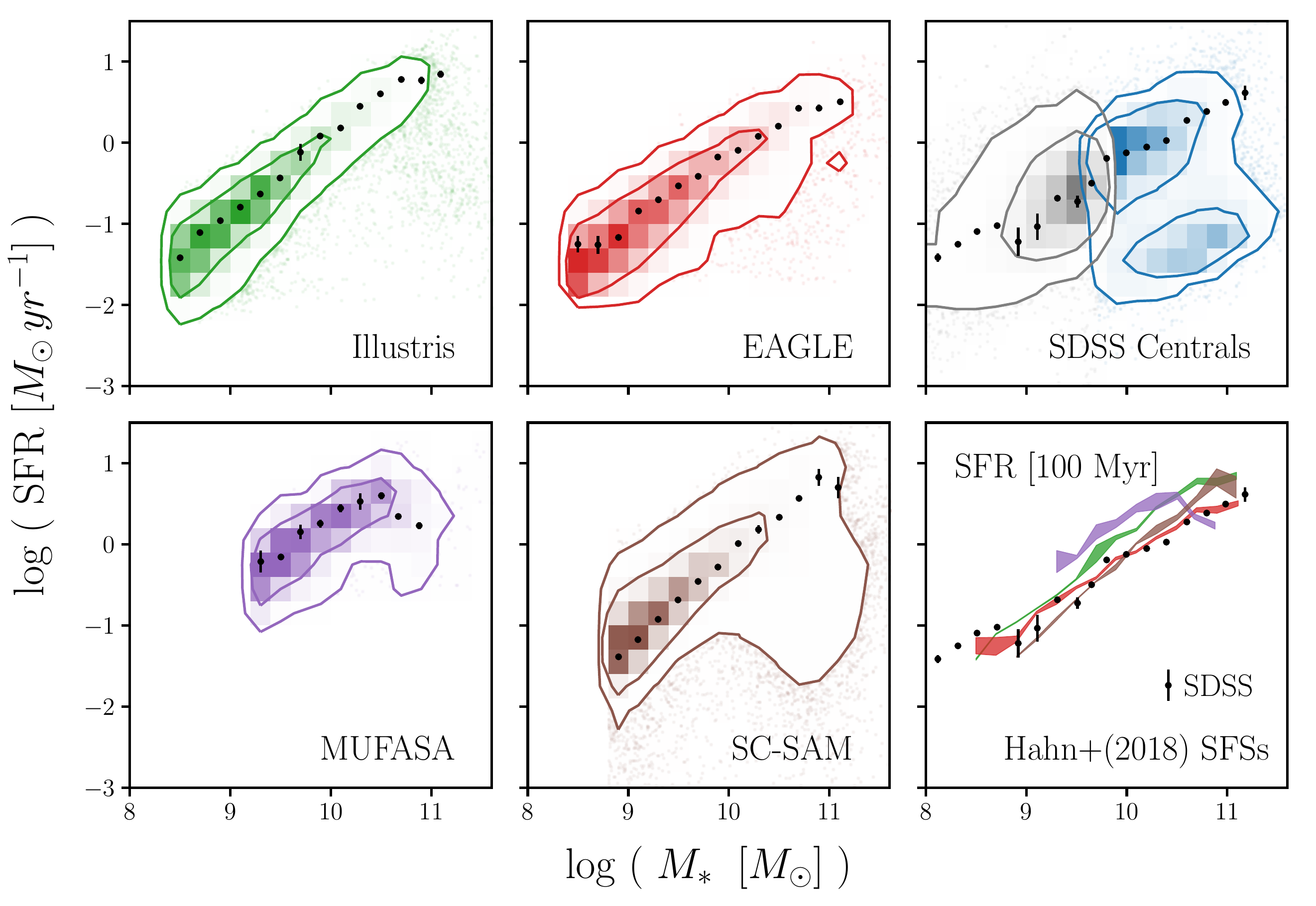} 
    \caption{Same as Figure~\ref{fig:sfmsfit_inst} but for $100\,\mathrm{Myr}$ SFR. 
    As in Figure~\ref{fig:sfmsfit_inst}, \emph{the SFSs of the simulations 
    have significantly different slopes and vary in amplitude by
    up to ${\sim}0.7\,\mathrm{dex}$, factor of ${\sim}5$}.}\label{fig:sfmsfit_100myr}
\end{center}
\end{figure*}

\begin{figure}
\begin{center}
\includegraphics[width = 0.48\textwidth]{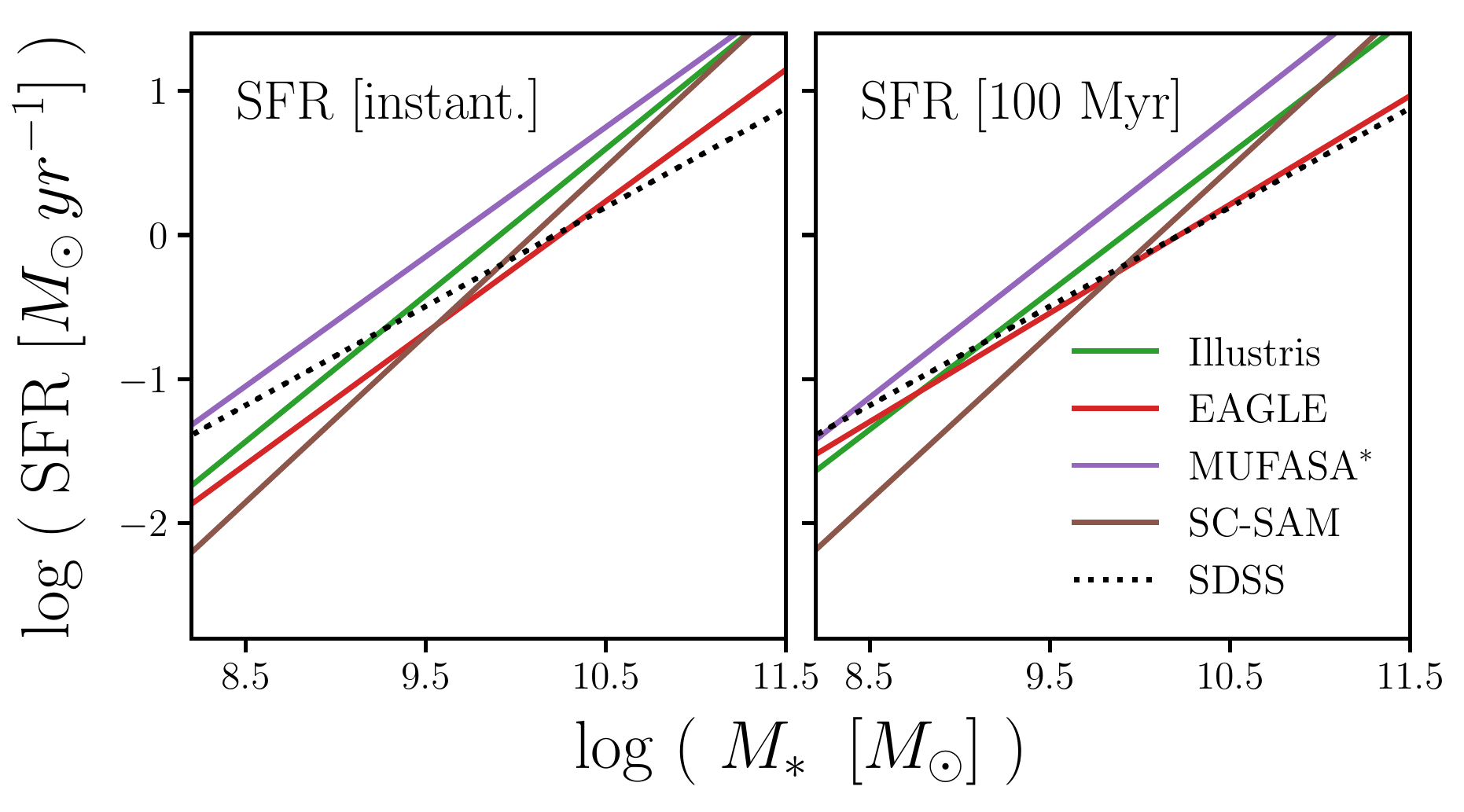} 
\caption{The power-law fits to the SFSs of the Illustris (green), 
    EAGLE (red), {\sc Mufasa} (purple), and SC-SAM (brown) simulations
   	highlight the significant differences in the slopes of the SFSs.
    We use instantaneous SFR and $100\,\mathrm{Myr}$ SFR in the left
    and right panels respectively. For reference, we include the 
    fit to the SDSS SFS (black dotted). We list the best-fit parameters 
    in Table~\ref{tab:sfms_powerlaw}. For a consistent comparison, we 
    fit the {\sc Mufasa} SFS below $\log\,M_* < 10.5$, due to its high stellar mass turnover.} 
    \label{fig:sfmsfit_powerlaw}
\end{center}
\end{figure}

\section{Results} \label{sec:results}
\subsection{SFS of simulated galaxies} \label{sec:sfs}
Now using our GMM based method from above, 
we can identify the SFSs of the simulated central galaxies from 
Section~\ref{sec:galsims}. We present the best-fit SFSs of the 
simulated galaxies from the Illustris, EAGLE, {\sc Mufasa}, and SC-SAM 
simulations for the instantaneous and $100\,\mathrm{Myr}$ SFR 
timescales in Figures~\ref{fig:sfmsfit_inst} and~\ref{fig:sfmsfit_100myr}, 
respectively. In each simulation, for both SFR timescales, the best-fit SFS is 
in good agreement with the underlying SFR-$M_*$ distribution as described 
by the contour and 2D histogram. However, when we compare the best-fit SFSs 
of the simulations to each another, \emph{we find that they have significantly 
different slopes and their amplitudes vary by up to 
$\,{\sim}0.7\,\mathrm{dex}$ (factor of ${\sim}5$) 
for both the instantaneous and \hunmyr~SFR timescales} 
(bottom right panels of Figures~\ref{fig:sfmsfit_inst} and~\ref{fig:sfmsfit_100myr}).

The uncertainties for the best-fit SFSs in Figures~\ref{fig:sfmsfit_inst} 
and~\ref{fig:sfmsfit_100myr} are derived from bootstrap resampling~\citep{efron1979} 
in each stellar mass bin. These uncertainties do not account for cosmic 
variance. 
Also, they correspond to the uncertainties of the means of the SFS GMM component, 
which is only one of the parameters in the GMM, and do \emph{not} account for 
the correlations other parameters of the GMM in Eq.~\ref{eq:gmm}. 
Our SFS uncertainties are estimated 
similarly to the cluster red sequence fits in~\cite{hao2009}, which use an 
``error-corrected'' GMM that involves bootstrap resampling. \cite{hao2009}, 
however, use their method to estimate the mean of their GMM component, 
rather than to estimate its uncertainty.

Using the SFSs we identified, we can now parameterize it to 
some functional form as often done in the literature --- \emph{e.g.} 
power-law~\citep{speagle2014} or broken power-law~\citep{lee2015}. With 
little evidence of a turnover in the SFS in {\em most} of simulations, 
we fit a power-law of the form 
\begin{equation} \label{eq:powerlaw}
\log\,\mathrm{SFR}_\mathrm{MS} = m\,(\log\,M_* - 10.5) + b
\end{equation}
to the SFSs in Figure~\ref{fig:sfmsfit_powerlaw}. Unlike the SFS of 
other simulations, the SFSs for {\sc Mufasa} have a significant 
turnover at $M_*{\sim}10^{10.5}M_\sun$. This turnover is \emph{not} 
caused by  misidentification of the SFS or some systematic effect in the 
GMM fitting. Instead, the turnover is due to the halo mass 
dependent quenching prescription in {\sc Mufasa} (Section~\ref{sec:mufasa}), 
which causes a sharper cut-off in the SFS, unlike the other more 
self-consistent AGN feedback models. We focus on the power-law portion 
of the {\sc Mufasa} SFS and fit Eq.~\ref{eq:powerlaw} below the turnover 
($M_*{<}10^{10.5} M_\sun$). 

The best-fit (least squares) power-law parameters (Table~\ref{tab:sfms_powerlaw} 
and Figure~\ref{fig:sfmsfit_powerlaw}) highlight the significant 
differences in the slope of the SFSs. Among our simulations, $m$ ranges 
from sub-linear in {\sc Mufasa} ($0.75$) to super-linear in SC-SAM ($1.17$).  
Various sub-grid models (\emph{e.g.} ISM, star  formation, stellar and 
AGN feedback) can influence the slope and normalization of SFSs in the 
simulations. To resolve the underlying cause behind the difference in SFSs, 
would require a detailed comparison of the different sub-grid parameters 
and prescriptions. While such a comparison is beyond the scope of this 
paper, the discrepancies we find in the SFSs provide constraints on 
galaxy formation models. Furthermore,
although a detailed comparison with observations is complicated by the 
differences in how SFR is defined in simulations versus observations, 
we include in Figure~\ref{fig:sfmsfit_powerlaw} the power-law fit to the 
SFS of the SDSS central galaxies (black dotted). Compared to the SFSs of the 
simulations, SFS in SDSS has a significantly a lower slope: $m=0.69$. As a 
result, the SFSs of the simulations are scattered around the SDSS SFS 
below $M_*{\sim}10^{10} M_\sun$, but have higher amplitudes than the SDSS SFS 
above $M_*{\sim}10^{10} M_\sun$. This is also apparent in the bottom right 
panel comparisons of Figures~\ref{fig:sfmsfit_inst} and~\ref{fig:sfmsfit_100myr}. 


The differences we find among the SFSs of the simulations, also propagate to their 
cosmic star formation densities. Cosmic star formation density roughly corresponds to the total 
star formation in the SFS weighted by the stellar mass function (SMF). For Illustris,
EAGLE, {\sc Mufasa}, and SC-SAM, respectively, we find total cosmic star formation 
densities (including satellites) of 
$10^{-1.66}, 10^{-2.22}, 10^{-1.87}$, and $10^{-1.94}\,M_\sun \mathrm{yr}^{-1} \mathrm{Mpc}^{-3}$ 
using instantaneous SFRs and similarly 
$10^{-1.68}$, $10^{-2.20}$, $10^{-1.91}$, and $10^{-1.94}\,M_\sun \mathrm{yr}^{-1} \mathrm{Mpc}^{-3}$
using $100\,\mathrm{Myr}$ SFRs. The rank order of the densities is different than that of the
SFSs due to differences in the SMFs. Although these values are 
roughly within the uncertainties of observations~\citep{madau2014}, 
the difference in the star formation density between Illustris and 
EAGLE, for example, is greater than $0.5\,\mathrm{dex}$--- more 
than factor of 3.

\begin{figure}
\begin{center}
\includegraphics[width=0.48\textwidth]{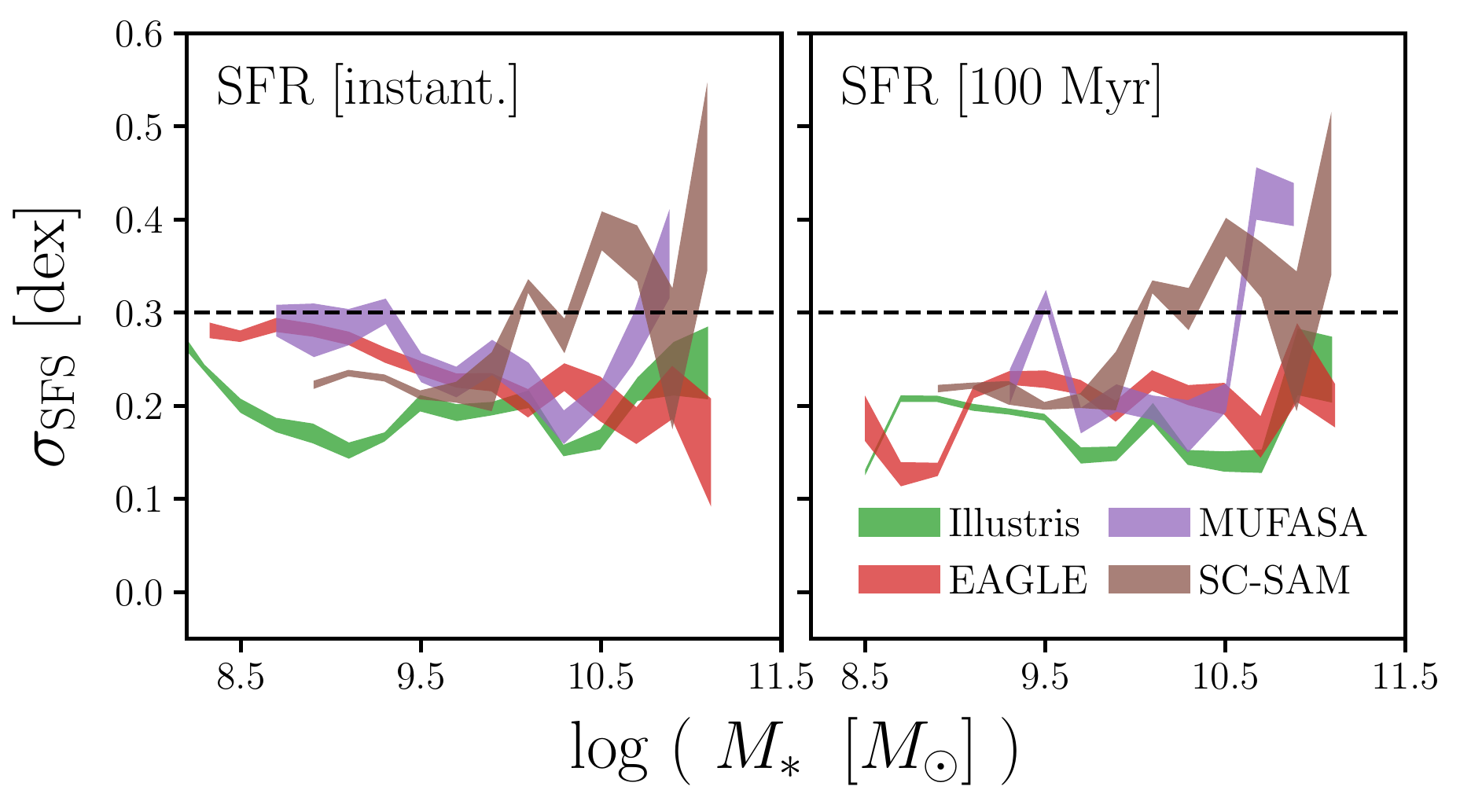}
\caption{The width of the SFS, $\sigma_\mathrm{SFS}$, for the simulated 
    central galaxies from Illustris, EAGLE, {\sc Mufasa}, and SC-SAM 
    (green, red, purple, and brown respectively). The uncertainties are 
    estimated using bootstrap resampling in the same way as the SFS uncertainties. 
    The SFS widths in the simulations have little stellar mass 
    dependence and, adding observational measurement errors in SFR, 
    they are roughly consistent with ${\sim}0.3\,\mathrm{dex}$ 
    from observations (black dashed).} \label{fig:sfms_width}
\end{center}
\end{figure}

In addition to its position, $\mu_\mathrm{SFS}$, the SFS GMM component 
is also described by $\sigma_\mathrm{SFS}$ --- the width of the SFS. 
Using $\sigma_\mathrm{SFS}$ derived from the GMM fitting, 
we can compare the width of the SFS among the simulations 
(Figure~\ref{fig:sfms_width}). The uncertainties for the widths are 
calculated through bootstrap resampling in the same way as the SFS 
uncertainties. Overall, we find little stellar mass 
dependence in $\sigma_\mathrm{SFS}$ for the simulations. For Illustris, 
EAGLE, {\sc Mufasa}, and SC-SAM we respectively find 
$\sigma_\mathrm{SFS}{\sim}0.20, 0.26, 0.25$, and $0.24\,\mathrm{dex}$ 
for instantaneous SFR and
$\sigma_\mathrm{SFS}{\sim}0.18, 0.20, 0.25$, and $0.23\,\mathrm{dex}$
for $100\,\mathrm{Myr}$ SFR (Table~\ref{tab:sfms_powerlaw}). Although 
we do not explicitly include the width of the SDSS SFS GMM component 
due to inconsistencies in the SFRs~(Section~\ref{sec:obvs}), these $\sigma_\mathrm{SFS}$ 
are narrower than the ${\sim}0.3\,\mathrm{dex}$ width measured in 
observations~\citep[\emph{e.g.}][]{daddi2007, noeske2007, salim2007, magdis2012, whitaker2012, speagle2014}. 
Observational errors, however, will bring the simulated values to closer 
agreement. Furthermore, the hydrodynamic simulations lack burstiness caused by 
clustered star formation (and thus feedback)~\citep{sparre2017a}. 
This unresolved variability will also bring the scatter closer to 
the observed width. 
We therefore conclude that \emph{the width of the SFS from the simulations 
are in agreement with the observed SFS width.}

\begin{table}
\caption{Power-law fit to the SFS of the simulated central galaxies from the
Illustris, EAGLE, {\sc Mufasa}, and SC-SAM simulations.} 
\begin{center}
\begin{tabular}{p{3cm}ccc} \toprule
\multicolumn{3}{c}{Star-Forming Sequence power-law fit} & width \\ [3pt]
\multicolumn{3}{c}{$\log\,\mathrm{SFR}_\mathrm{MS} = m\,(\log\,M_* - 10.5) + b$  } & \\ [3pt]
Simulation & $m$ & $b$ & $\sigma_\mathrm{SFS}$ [dex] \\ 
\hline
\multicolumn{4}{c}{Instantaneous SFR} \\
Illustris 			& 1.01 $\pm$ 0.004 & 0.59 $\pm$ 0.006 & 0.20 \\ 
EAGLE 				& 0.91 $\pm$ 0.006 & 0.23 $\pm$ 0.008 & 0.26 \\ 
{\sc Mufasa} 		& 0.75 $\pm$ 0.014 & 0.58 $\pm$ 0.011 & 0.25 \\ 
{\sc Mufasa}$^*$ 	& 0.89 $\pm$ 0.020 & 0.74 $\pm$ 0.020 &  \\ 
SC-SAM 				& 1.17 $\pm$ 0.008 & 0.48 $\pm$ 0.009 & 0.24 \\ 
\hline \hline
\multicolumn{4}{c}{$100\,\mathrm{Myr}$ SFR} \\
Illustris 			& 0.95 $\pm$ 0.006 & 0.55 $\pm$ 0.008 & 0.18 \\
EAGLE  				& 0.75 $\pm$ 0.010 & 0.21 $\pm$ 0.009 & 0.20 \\
{\sc Mufasa}		& 0.38 $\pm$ 0.023 & 0.36 $\pm$ 0.016 & 0.25 \\
{\sc Mufasa}$^*$ 	& 0.97 $\pm$ 0.050 & 0.83 $\pm$ 0.039 & \\ 
SC-SAM 				& 1.16 $\pm$ 0.008 & 0.47 $\pm$ 0.009 & 0.23\\ 
\hline
\hline \hspace{10pt}
SDSS 				& 0.69 $\pm$ 0.008 & 0.18 $\pm$ 0.007 \\ 
\hline
\end{tabular} \label{tab:sfms_powerlaw}
\end{center}
$^*$ power-law fit to the {\sc Mufasa} SFS below its turnover ($\log\,M_* {<}\,10.5$)
\end{table}

One factor that impacts the SFS we identify is the strict lower limit of the 
$\log\mathrm{SFR}$s caused by the resolution effects in the simulations. 
This is particularly evident in the $100\,\mathrm{Myr}$ SFR--$M_*$ 
relations of the hydrodynamic simulations of Figure~\ref{fig:sfrmstar} --- especially 
{\sc Mufasa}. As we describe in Section~\ref{sec:galsims}, the $100\,\mathrm{Myr}$ 
SFRs are  calculated using the ages of all star particles in a galaxy. For a galaxy to 
have star formation (\emph{i.e.} SFR $> 0$), it must \emph{at least} 
form one star particle over the last $100\,\mathrm{Myr}$. A single star particle 
forming over $100\,\mathrm{Myr}$ amounts to a SFR of 
${\sim}0.02\ M_{\sun} \mathrm{yr}^{-1}$ for Illustris and EAGLE and
${\sim}0.2\ M_{\sun} \mathrm{yr}^{-1}$ for {\sc Mufasa}. This resolution limit, ultimately 
impacts the SFS at $M_*{<}10^{8.4}$, $10^{8.4}$, and 
$10^{9.2}M_\sun$ for Illustris, EAGLE, and {\sc Mufasa} respectively (see Appendix~\ref{app:zerosfr}). 

Using our method for identifying the SFS, we are able to 
conduct a consistent data-driven comparison of the SFSs of simulated 
central galaxies from the Illustris, EAGLE, {\sc Mufasa}, and SC-SAM. From 
this comparison, we find that the amplitudes of the SFSs differ from one 
another by up to ${\sim}0.7\,\mathrm{dex}$, factor of ${\sim}5$, 
with significantly different slopes. Furthermore, despite these differences, the SFSs of 
the simulations have similar widths, consistent with observations. 

\begin{figure*}
\begin{center}
\includegraphics[width=0.95\textwidth]{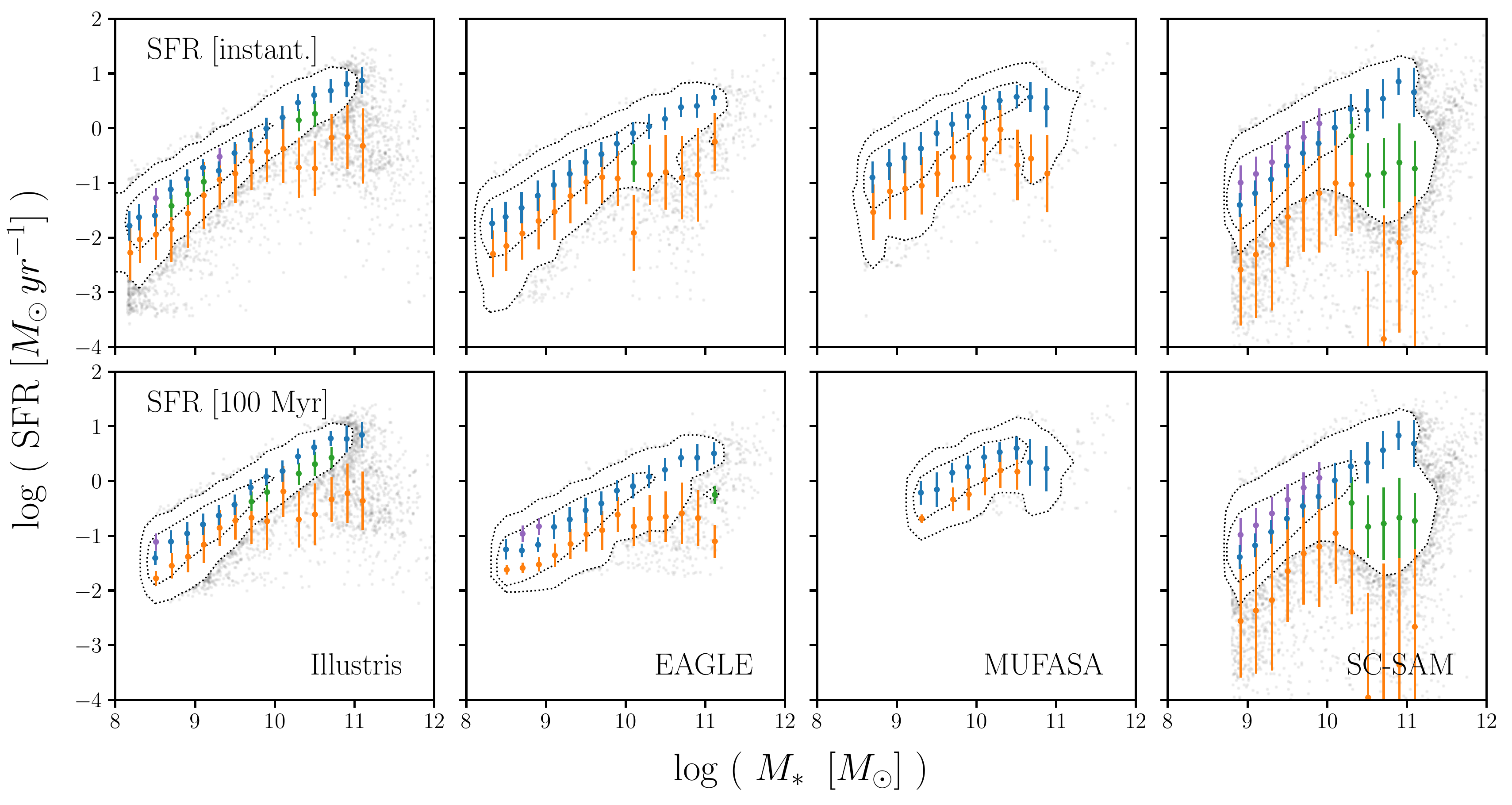} 
\caption{Components of the best-fit GMM for the SFR-$M_*$ relations of central 
    galaxies in the Illustris, EAGLE, {\sc Mufasa}, and 
    SC-SAM simulations (left to right). The top and bottom panels use instantaneous 
    SFRs and $100\,\mathrm{Myr}$ SFRs respectively. In each $\log M_*$ bin, we mark 
    the SFS component in blue, the low SF component in orange, the intermediate SF 
    component in green, and the component above the SFS in purple. These components 
    \emph{loosely} correspond to the star-forming, quiescent, transitioning, and 
    star-burst subpopulations. The hydrodynamic simulations have similar 
    subpopulations dominated by the SFS and low SF components. Meanwhile in the 
    SC-SAM, the GMM components reveal broad low SF components that extends out to 
    $\mathrm{SFR} < 10^{-4}M_\sun \mathrm{yr}^{-1}$, prominent intermediate components at 
    $M_* \gtrsim 10^{10}M_\sun$, and components above the SFS at $M_* \lesssim 10^{10}M_\sun$.} \label{fig:sfmsfit_comps}
\end{center}
\end{figure*}

\begin{figure*}
\begin{center}
\includegraphics[width=\textwidth]{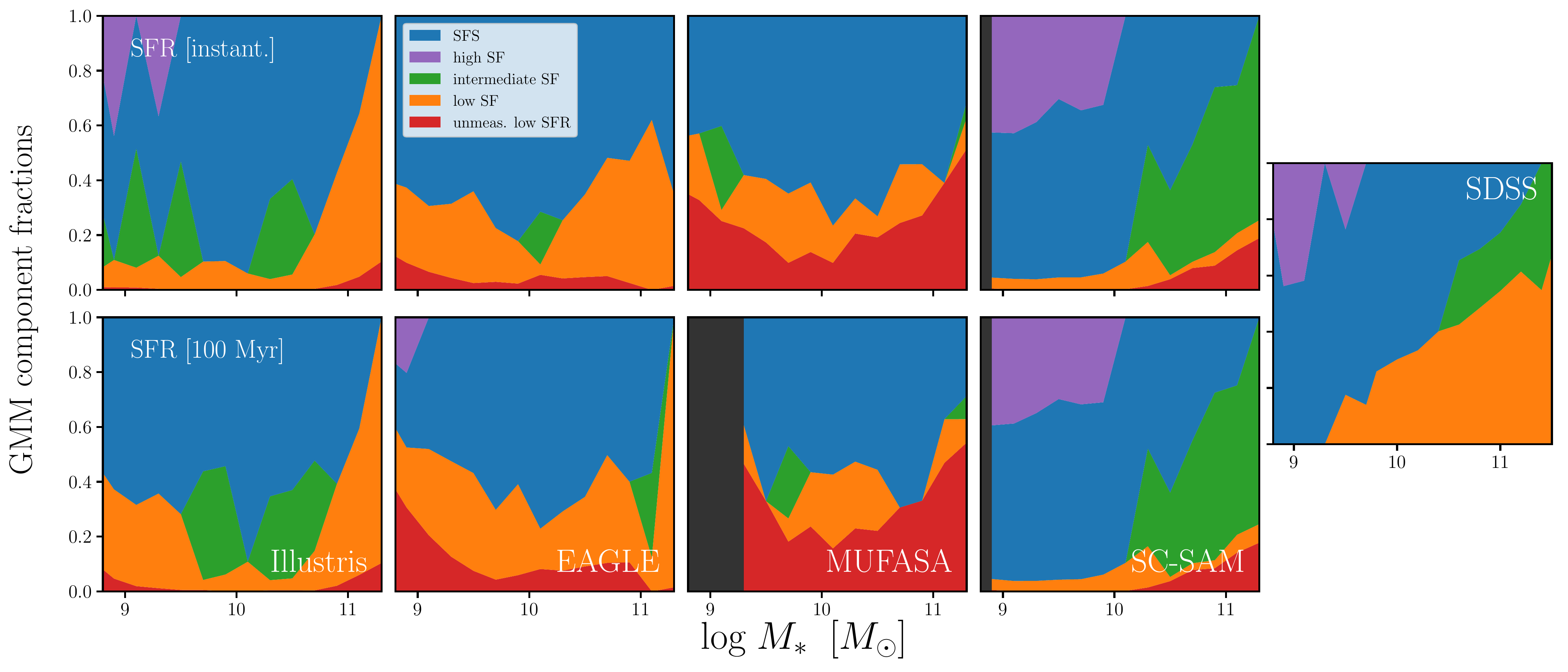} 
\caption{Fractional contributions, $\pi_i$, of the best-fit GMM components of 
    the central galaxies in Illustris, EAGLE, {\sc Mufasa}, and SC-SAM (left 
    to right). We highlight the SFS component in blue, the low SF component 
    in orange, galaxies with unmeasurably low SFR in red, the intermediate SF 
    components in green, and the high SF components in purple. We shade the 
    regions below the stellar mass limit set by resolution effects in black 
    (Appendix~\ref{app:zerosfr}). For reference, we include $\pi_i$ of the 
    observed SDSS centrals in the rightmost panel. Unlike SDSS or the SC-SAM, 
    we do not find significant high SF components at low $M_*$ in the hydrodynamic
    simulations. Furthermore, treating the components below the SFS as quiescent,
    we find little $M_*$ dependence in the quiescent fraction at $M_* < 10^{11}M_\sun$
    unlike observations. In fact, in all of the simulations, we find a significant 
    fraction of quiescent central galaxies at $M_* \lesssim 10^9 M_\sun$ contrary to observations.} \label{fig:kandinsky}
\end{center}
\end{figure*}

\subsection{Beyond the SFS of Simulated Galaxies} \label{sec:beyondsfms}
So far we have focused solely on the SFSs of the simulated galaxies --- \emph{i.e.} 
$\bm{\theta}_\mathrm{SFS} = \{\mu_\mathrm{SFS}, \sigma_\mathrm{SFS} \}$ 
in Eq.~\ref{eq:gmm}. Our GMM method, however, also determines 
$\bm{\theta}_i$ of components other than the SFS. These GMM components 
provide extra features to compare the simulated galaxy samples and also 
offer interesting insights into the different subpopulations in the simulated 
galaxy samples. When we examine $\bm{\theta}_i$ of all components from 
our fitting for the simulated galaxies, we find they loosely correspond 
to galaxy subpopulations typically referred to as quiescent, transitioning, 
and star-burst (Figure~\ref{fig:sfmsfit_comps}). To avoid over-interpreting 
this correspondence, we refer to the GMM component with the lowest SFR as 
``low SF'' component, the component with SFR in between the SFS and the 
low SF component as the ``intermediate SF'' component, and finally the 
component with higher SFR than the SFS component as the ``high SF'' component. 
At a given stellar mass bin, our GMM fits are restricted to $k\leq3$; hence, 
the four different components come from different stellar mass bins. 
In Figure~\ref{fig:sfmsfit_comps}, we mark the SFS, low SF, intermediate SF, 
and high SF in blue, orange, green, and purple respectively.

Examining the GMM components of the hydrodynamic simulations in 
Figure~\ref{fig:sfmsfit_comps}, we find that a few $\log M_*$ bins have 
intermediate SF components in Illustris at $10^9 M_\sun < M_* < 10^{11}M_\sun$.
Also a few of the lowest $\log M_*$ bins in Illustris and EAGLE have high SF
components for the $100\,\mathrm{Myr}$ SFRs. Besides these few 
bins, however, the central galaxies from the hydrodynamic simulations are
dominated by the SFS and low SF components. Furthermore, 
throughout the stellar mass ranges of the simulations, the low SF 
components in each of these simulations have relatively constant widths and 
lie ${\sim}1\,\mathrm{dex}$ below the SFS components.

Unlike the hydrodynamic simulations, however, the low SF components in the 
SC-SAM span out to $\log\,\mathrm{SFR}{=}{-}4\ M_\sun \mathrm{yr}^{-1}$. Furthermore, the intermediate 
and high SF components are much more prominent in the SC-SAM centrals. 
At low stellar masses ($M_* \lesssim 10^{10}M_\sun$) every $\log M_*$ 
bin has a high SF component. The $\log\,\mathrm{SSFR}$ distributions in these 
bins have extended tails on the higher SFR side of the SFS. Our GMM 
method, thus, fits high SF components in these $\log M_*$ bins (bottom left 
and center panels of Figures~\ref{fig:pssfr_gmm_inst} and~\ref{fig:pssfr_gmm_100myr}). 
These high SF components and the extended range of low SF components are likely 
caused by the re-accretion prescription of the SAM (Section 2.7 of~\citealt{somerville2008a}).
A fraction of gas ejected from halos (\emph{e.g.} from supernovae) is kept in 
a reservoir, which re-collapses into the halos at a later time and becomes available 
again for cooling. The rate of this re-accretion depends on the mass of ejected gas, 
the dynamical time of the halo, and a free parameter degenerate with supernovae 
feedback parameters. This prescription results in bursty 
star formation in the SC-SAM galaxies and causes the extended low SF components and 
the high SF components. 


At high stellar masses ($M_* \gtrsim 10^{10}M_\sun$) every $\log M_*$ bin in the 
SC-SAM has an intermediate component. While the $\log\,\mathrm{SSFR}$ 
distributions in the bottom right panels of Figures~\ref{fig:pssfr_gmm_inst} 
and~\ref{fig:pssfr_gmm_100myr} and the BIC values illustrate the benefit of the 
GMM with an intermediate SF component, these are accentuated by the broader 
distribution of the low SF population. Despite these differences between the 
hydrodynamic simulations and the SC-SAM, all of the simulations have a low SF 
component throughout their stellar mass 
range, even at $M_* < 10^9M_\sun$. We discuss these low $M_*$ low SF galaxies 
in further detail later in this section.

Another set of parameters we infer from our GMM fitting is the weight of the 
GMM components: $\pi_i$ in Eq.~\ref{eq:gmm}. These weights correspond to 
the fractional contribution of the different subpopulations. For example, the 
weight of the low SF component loosely corresponds to the quiescent 
fraction~\citep[\emph{e.g.}][]{borch2006, bundy2006, iovino2010, geha2012, hahn2015}. 
In Figure~\ref{fig:kandinsky}, we present the fractional contribution of the 
components from our best-fit GMM, as a function of stellar mass: SFS (blue), 
low SF (orange),  intermediate SF (green), and high SF (purple). We also include
the fractional contribution of galaxies with unmesurably low SFRs (red; 
see Section~\ref{sec:galsims}). The $\pi_i$ have 
uncertainties, estimated from bootstrap resampling, on the order of ${\sim}0.1$.

For every simulation, a significant fraction of galaxies have unmeasurably 
low SFRs. In hydrodynamic simulations, a galaxy with unmeasureably low SFR 
can have an SFR below the resolution limit, or have a ``true'' $\mathrm{SFR}{=}0$ 
on the measured timescales (Appendix~\ref{app:zerosfr}). For the SC-SAM, we 
consider the SFR unmeasurably low when $\log \mathrm{SFR}< {-}4\ M_\sun \mathrm{yr}^{-1}$. 
Therefore, in both hydrodynamic simulations and the SAM, galaxies with 
unmeasurably low SFR can be considered quiescent. 
Moreover, we confirm that SFR resolution does not significantly 
impact the fraction contributions of Figures~\ref{fig:kandinsky} 
(see Appendix~\ref{app:zerosfr} and Figure~\ref{fig:mlim_fcomp}).

The fractional contributions of the GMM components in Figures~\ref{fig:kandinsky} 
reveal significant disagreements between the simulated galaxies and 
trends established from observations--- especially the hydrodynamic 
simulations. 
For instance, in the hydrodynamic simulations we do not find significant 
high SF components at low $M_*$, unlike in SDSS or SC-SAM. The few $M_*$ 
bins with fractional contributions from high SF components have large 
bootstrap uncertainties (${\sim}0.2$). 
Furthermore, if we treat the components below the SFS as quiescent 
(green, orange and red in Figure~\ref{fig:kandinsky}),
\emph{we find little stellar mass dependence in the quiescent fraction of the 
hydrodynamic simulations, unlike the quiescent fraction measurements 
of isolated SDSS galaxies}~\citep{baldry2006,peng2010,hahn2015}. 
Meanwhile, at $M_*{>}10^9M_\sun$ the SC-SAM is roughly consistent with 
SDSS (rightmost panel) and in agreement with previous 
SC-SAM quiescent fraction comparisons to observations~\citep{brennan2015,brennan2017,pandya2017}.

Furthermore, for some of the hydrodynamic simulations in 
Figure~\ref{fig:kandinsky} (Illustris, EAGLE, and {\sc Mufasa} with 
$100\,\mathrm{Myr}$ SFRs and EAGLE, and {\sc Mufasa} instantaneous SFRs)
we find surprisingly high quiescent fractions (${\sim}0.4$) at low masses in stark 
contrast with observations~\citep{baldry2006,peng2010,hahn2015}. In fact, 
\emph{all the simulations, even the SC-SAM, have non-negligible (${\gtrsim}10\%$) 
quiescent fraction at $M_*{<}10^9 M_\sun$ contrary to the $M_*$ lower bound of 
${\sim}10^9M_\sun$ for isolated/central quiescent galaxies 
we observe in SDSS and established in the literature~\citep[\emph{e.g.}][]{geha2012}.}

One possible explanation for the significant fraction of low SFR galaxies
at low $M_*$ in the hydrodynamic simulations is misclassification of 
``splashback'' (or ``blacksplash'' or ``ejected'') galaxies as centrals. 
Splashback galaxies are satellite galaxies that have orbited outside 
the virial radii of its host halo after having passed through 
it~\citep[\emph{e.g.}][]{mamon2004,gill2005,wang2009a,wetzel2014}.
The SC-SAM is {\em not} subject to this misclassification 
because subhalos are not tracked after mergers, so by construction the 
model does not have splashbacks. To test whether splashbacks 
impact our results for the hydrodynamic simulations, we adjust our central
galaxy selection criteria in Section~\ref{sec:central} to exclude 
any centrals with a more massive halo within three virial radii of it. 
When we use this stricter central classification and measure the
SFS and other GMM components, we find \emph{no} significant change to 
the SFS fits or the fractional contributions of the GMM components. 
We also find no significant changes to our results when we restrict 
the selection to galaxies with no ``luminous'' neighbors within 
$1.5\,\mathrm{Mpc}/h$ --- analogous to the~\cite{geha2012} criteria.
We therefore conclude that the significant fraction of low SFR and 
low $M_*$ galaxies is not caused by misclassification of centrals.

Another possible explanation for the abundance of low SFR galaxies
at low $M_*$, is that the hydrodynamic simulations have insufficient
resolution for galaxies with $M_*< 10^9M_\sun$. Low $M_*$ galaxies 
in reality, may have star-forming clumps with masses lower than the 
baryonic particle mass. Such star formation will not be captured by 
the simulations~\citep{sparre2017a}. We test whether our results are 
impacted by the resolution limit using a higher resolution box ($8\times$ 
higher baryon mass resolution) for EAGLE. When we measure the fractional 
contributions of the GMM components for the higher resolution EAGLE 
simulation, we confirm the abundance of low SFR galaxies at 
$M_*<10^9M_\sun$.  

Taking a step back, we emphasize that this discrepancy between the 
simulations and observations must be taken with a grain of salt and
our comparison is not an apples-to-apples comparison. For instance,
in \cite{geha2012} low SF/quiescent galaxies are classified based on 
a $H\alpha$ emission and $D_n 4000$ criteria --- different than in 
the simulations. Even the central (isolation in~\citealt{geha2012}) 
criteria, in detail, is different than the analogous criteria 
above. More broadly, the comparisons we present in this paper are 
among simulations and therefore are based on ``theoretical'' predictions 
of galaxy properties. Many factors make it difficult 
to robustly extend this comparison to observations. 

For example, SFRs and $M_*$, the galaxy properties considered in this paper, 
in simulations can be directly measured either using star or gas particles in 
the simulations. In observations, even the SFR alone is estimated from SFR indicators 
such as $H\alpha$ flux, $D_n 4000$, or UV brightness and dust absorption measurements. 
While they serve as estimates of the SFRs, as \cite{speagle2014} find even 
for the same SDSS galaxies, different SFR indicators can produce large 
discrepancies in the slope and amplitude of the SFS. Furthermore, a 
consistent comparison to observations requires a thorough understanding of 
the selection effects that come with the observed galaxy sample. These 
effects are difficult to propagate into SFR and $M_*$ space of simulations. 


Therefore, while we note some of the differences in 
Figures~\ref{fig:sfmsfit_inst},~\ref{fig:sfmsfit_100myr},~\ref{fig:sfmsfit_powerlaw}, 
and~\ref{fig:kandinsky}, between the simulations and observations,  
we reserve a more detailed comparison to the next paper in our 
series: Starkenburg et al. in prep. In this next paper, instead of
comparing the ``theoretical'' galaxy properties, we forward model galaxy 
spectra and photometry of simulated galaxies using their star formation 
histories, make observationally motivated measurements of SFR and $M_*$ on
the synthetic spectra and photometry, and conduct a quantitative, 
apples-to-apples, comparison of the simulations to observations.

In this section, we demonstrate that our method for identifying the SFS
provides additional features besides the SFSs, to compare different galaxy 
samples. These extra components offer insights into the 
distinct galaxy subpopulations of the simulations. Based on the non-SFS 
components/populations, we find that the hydrodynamic simulations are 
similarly dominated by the SFS and low SF components, while the SC-SAM 
predicts substantial fractions of high and intermediate SF components. 
Moreover, we find that all of the simulations have a significant
fraction of low SFR central galaxies at $M_*\,{\lesssim}\,10^9M_\sun$, 
contrary to observations. Furthermore, the hydrodynamic simulations, at 
even $M_*\,{\lesssim}\,10^{11}M_\sun$, do not reproduce the quiescent 
fractions from the literature or their stellar mass dependence. 

\section{Summary and Conclusions} \label{sec:summary}
The Star-Forming Sequence provides a key feature in galaxy property 
space to consistently compare 
galaxy populations in simulations and observations.
Such comparisons are crucial for validating our theories of
galaxy formation and evolution. However, they face two main challenges: 
the lack of a consistent data-driven method for identifying the SFS and 
the discrepancies in methodology for deriving galaxy properties such as 
SFR and $M_*$. In this paper, we address the former by presenting 
a flexible data-driven method for identifying the SFS. 

Our method takes advantage of Gaussian mixture models to fit the SFR
distributions in stellar mass bins and Bayesian Information Criteria 
for model selection. This data-driven approach allows us to robustly 
fit the SFR-$M_*$ relation of galaxy populations and identify the SFS, 
while relaxing many of the assumptions and hard cuts that go into 
other methods. Furthermore, it allows us to identify the SFS over a wide 
range of star formation and stellar masses down to 
$M_*{\sim}10^{8}M_\sun$. Finally, our method also allows us to identify 
subpopulations of galaxies, beyond the SFS, that correspond to the 
quiescent, transitioning, and star-burst galaxy populations. 

Next we apply our method to the central galaxies of the Illustris, EAGLE, 
and {\sc Mufasa} hydrodynamic simulations and the Santa 
Cruz Semi-Analytic Model. The central galaxies are identified in the  
simulations using the \cite{tinker2011} group 
finder and have \emph{consistently} derived $M_*$ and SFRs on instantaneous 
and $100\,\mathrm{Myr}$ timescales. For reference, we also apply our 
method to central galaxies from SDSS observations. Comparing the resulting 
SFSs and other components from the simulations and observations, we find the following:

\begin{itemize}
\item The identified SFSs of Illustris, EAGLE, {\sc Mufasa}, and SC-SAM 
vary by up to ${\sim}0.7\,\mathrm{dex}$ (factor of ${\sim}5$) 
and have significantly different slopes over the stellar mass
range $10^{8.5} M_\sun < M_* < 10^{11} M_\sun$ with little mass
dependence in the discrepancies. Meanwhile the width of the SFSs 
are consistent with one another and in agreement with the 
$\sim 0.3\,\mathrm{dex}$ width from observations.

\item From the best-fit GMMs, we find that the hydrodynamic simulations are 
mainly dominated by the SFS and low SF (quiescent) components. Meanwhile,
the SC-SAM is composed of a substantial fraction of galaxies between the 
SFS and low SF components at high masses ($M_* > 10^{10}M_\sun$) and above 
the SFS at low masses ($M_* < 10^{10}M_\sun$), likely due to its 
re-accretion prescription. 

\item The quiescent fractions of the hydrodynamic simulations, estimated 
from the components of the best-fit GMMs and galaxies with unmeasurably 
low SFR, have little stellar mass dependence and are inconsistent 
with the SC-SAM as well as with observations. Moreover, in all of 
the simulations, we find an abundance of low mass ($M_*{<}\,10^9M_\sun$)
quiescent central galaxies, which we do not find in SDSS or the literature .
\end{itemize}

With a consistent treatment of the simulations and our method
for identifying their SFSs and other subpopulations,
we demonstrate significant differences in the  
central galaxy populations of Illustris, EAGLE,  {\sc Mufasa}, and SC-SAM. 
Although we refrain from a detailed 
comparison with observations, we also find significant differences 
between the simulations and established trends in observations. These
discrepancies, which previous comparisons failed to identify, underscore 
the importance of a consistent data-driven approach for accurately comparing 
galaxy populations.

Furthermore these results illustrate how differences in the 
sub-grid physics of the simulations propagate into significant differences  
in the properties of their galaxy populations. Extending our approach of 
a consistent data-driven comparison, to observations, we can test the 
subgrid physics of simulations and derive strong constraints on our galaxy 
formation models. This is exactly what we will present in the subsequent 
paper of our series---Starkenburg et al. in prep.


\section*{Acknowledgements}
It is a pleasure to thank
	Melanie~Habouzit,
	Shirley~Ho, 
    John~Moustakas,
    and 
	Emmanuel~Schaan 
for valuable discussions and feedback. 
We also thank the Illustris collaboration and the Virgo Consortium for making 
their simulation data publicly available. The EAGLE simulations were performed 
using the DiRAC-2 facility at Durham, managed by the ICC, and the PRACE facility 
Curie based in France at TGCC, CEA, Bruy\`{e}res-le-Ch\^{a}tel.
This material is based upon work supported by the U.S. Department
of Energy, Office of Science, Office of High Energy Physics, under
contract No. DE-AC02-05CH11231. The IQ (Isolated \& Quiescent)-Collaboratory thanks the Flatiron Institute for hosting the collaboratory 
and its meetings. The Flatiron Institute is supported by the Simons Foundation.
\appendix
\counterwithin{figure}{section}

\section{Previous Comparisons of the Star-Forming Sequence} \label{app:literature}
Earlier SFS comparisons in the literature overall report agreement among 
simulations and observations at $z=0$~\citep[\emph{e.g.}][]{genel2014, somerville2015, sparre2015, schaye2015, bluck2016, dave2016}. 
This agreement is particularly evident in the comparison in 
\cite{somerville2015} (Figure 5). However, as \cite{somerville2015} note, 
the SFSs compiled in the comparison are derived inconsistently, with some 
applying a star-forming galaxy selection cut (\emph{e.g.} SSFR cut) and 
others not applying any cut. We demonstrate in this section that {\em inconsistency 
in measuring the SFS can produce misleading agreement among simulations.} 

\begin{figure*}
\begin{center}
\includegraphics[width=0.5\textwidth]{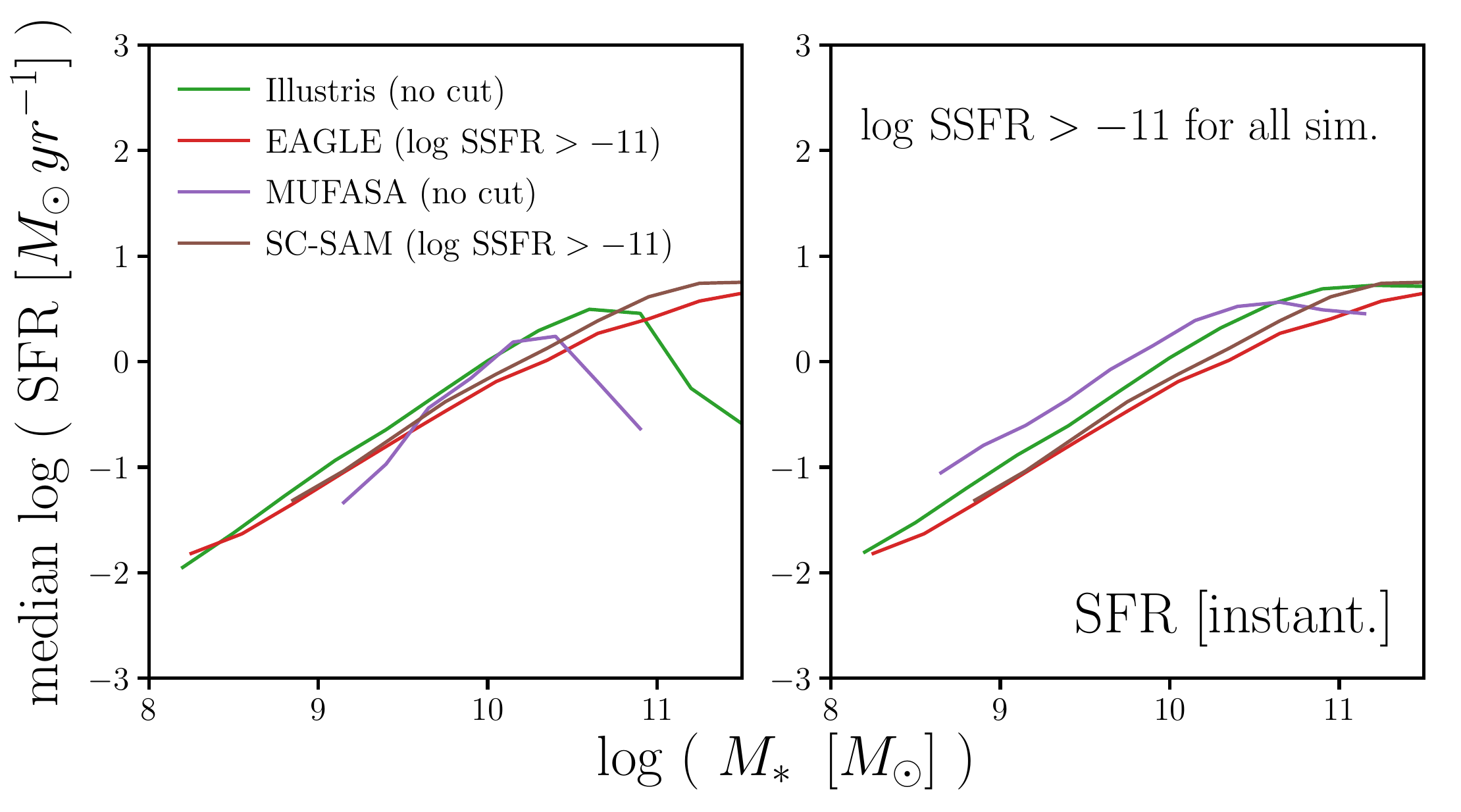} 
\caption{The SFSs of Illustris, EAGLE, {\sc Mufasa}, and SC-SAM central 
galaxies, where we measure the SFSs using different methods as in Figure 5 
of \cite{somerville2015} (left panel) and using the same method (right panel). 
In the left panel, we measure the SFSs by taking the median SFR in 
a $M_*$ bin with no selection cuts for Illustris and {\sc Mufasa}
and by taking the median SFR after a SSFR > $10^{-11}\, \mathrm{yr}^{-1}$ cut 
for EAGLE and SC-SAM. In the right panel, we measure the SFSs by taking 
the median SFR after a SSFR > $10^{-11}\,\mathrm{yr}^{-1}$ cut for all four simulations.
The difference between the two panels illustrate that \emph{the agreement 
found in the left panel, and similarly in \cite{somerville2015},
is mainly driven by the difference in methods used to measure SFSs.}
} 
\label{fig:likeSD14}
\end{center}
\end{figure*}

In the left panel of Figure~\ref{fig:likeSD14} we reproduce the SFS comparison of 
\cite{somerville2015} Figure 5 for the simulations in Section~\ref{sec:galsims}
using different methods for measuring the SFS. For Illustris and EAGLE, we apply
the same methods as the SFSs in \cite{somerville2015}: 
the median SFR in a $M_*$ bin with no selection cut for Illustris (green) and 
with a SSFR > $10^{-11}\, \mathrm{yr}^{-1}$ cut for EAGLE~\citep[red;][]{schaye2015}. 
{\sc Mufasa} and the current version of SC-SAM did not exist and were not 
included in \cite{somerville2015}. Since we are illustrating how inconsistent 
SFS measurements can result in misleading agreement, for {\sc Mufasa} and 
SC-SAM we measure the median SFR with no selection cut and with a SSFR > 
$10^{-11}\, \mathrm{yr}^{-1}$ cut, respectively. As in Figure 5 of \cite{somerville2015}, 
we find good agreement among the SFSs of the simulations. 

Instead of measuring the SFSs differently, if we measure the SFS by taking the
median SFR after a SSFR > $10^{-11}\, \mathrm{yr}^{-1}$ cut consistently for all the  
simulations, we find discrepancies in the SFSs on the order of ${\sim}0.5\,\mathrm{dex}$
(right panel of \ref{fig:likeSD14}). This illustrates that the agreement found in 
\cite{somerville2015} is driven in large part by the difference in methods used to 
measure SFSs. Furthermore, the difference between the Illustris and {\sc Mufasa} SFSs in 
the two panels illustrate how different SFS fitting methods, even when 
consistently applied, can increase the difference between SFSs. The difference 
between the Illustris and {\sc Mufasa} SFSs is significantly larger in the 
right panel when the SSFR > $10^{-11}\, \mathrm{yr}^{-1}$ cut is applied and 
galaxies with SFR below the resolution limit, which make up a larger fraction of the {\sc Mufasa} 
galaxies than the Illustris galaxies, are removed by the selection cut. 
This impact highlights the need for a data-driven method for identifying the 
SFS that better account for intrinsically different SFR--$M_*$ distributions 
in the simulations. Therefore, Figure~\ref{fig:likeSD14} highlights the impact 
of hard selection cuts and the importance of using a consistent data-driven 
methodology for measuring the SFS.

\section{Identifying the SFS using Gaussian Mixture Models} \label{app:gmm_pssfr}
\begin{figure*}
\begin{center}
\includegraphics[width=0.95\textwidth]{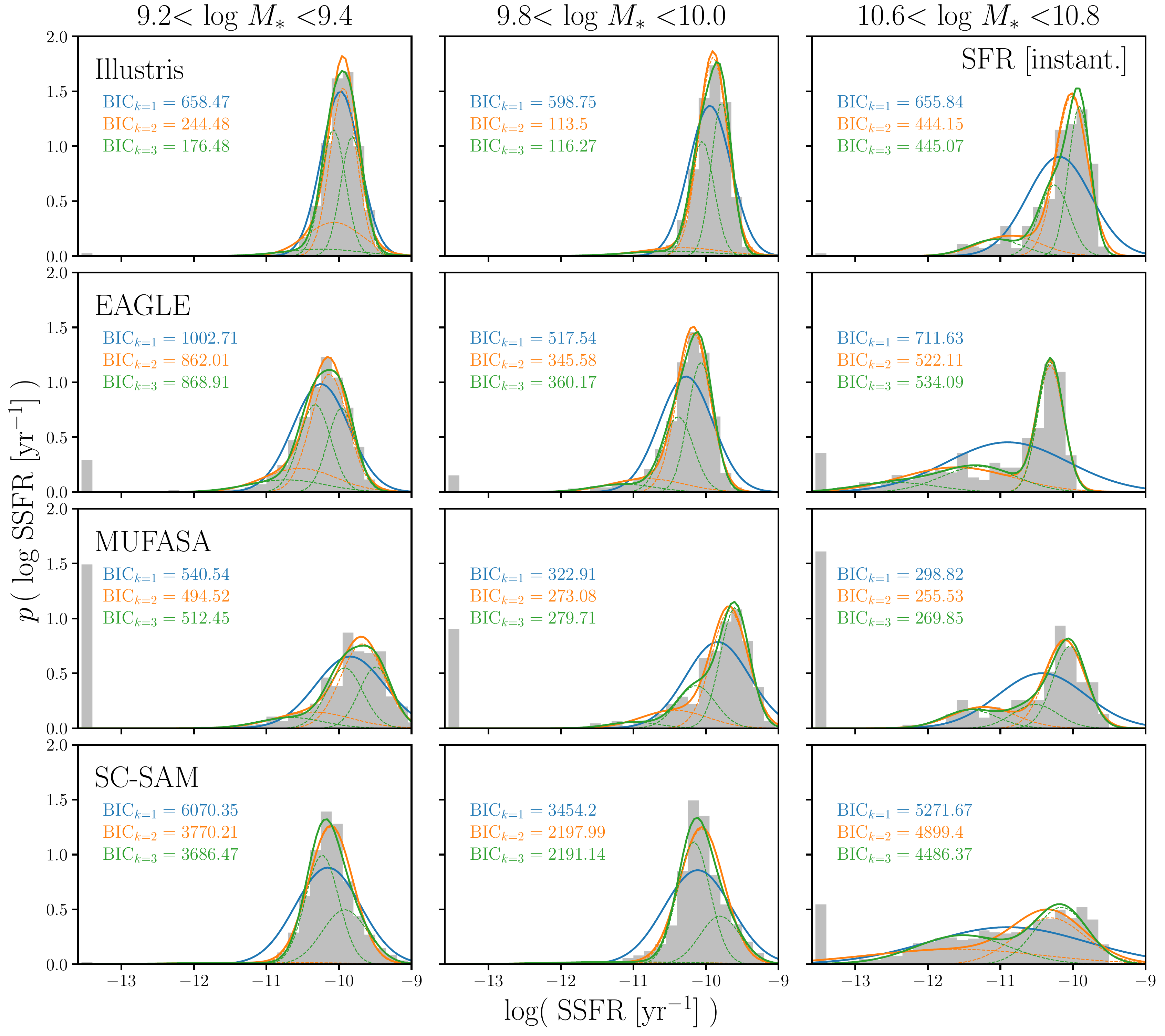} 
\caption{GMMs with $k=1, 2,$ and $3$ (blue, orange, and green) components fit
to the instantaneous SSFR distributions, $P(\log\,\mathrm{SSFR})$, of the 
Illustris, EAGLE, {\sc Mufasa}, and SC-SAM (top to bottom panels) centrals 
in three stellar mass bins: $[9.2, 9.4]$, $[9.8, 10.]$, and $[10.6, 10.8]$ 
(left to right). We represent galaxies with unmeasurably low SFR in 
$P(\log\,\mathrm{SSFR})$s with $\log\,\mathrm{SSFR} = -13.5$. 
For every GMM fit, we plot each component in dash lines 
and list their BICs in the same color. In our SFS fitting, we select the 
GMM with the lowest BIC as the best-fit. This provides a \emph{data-driven 
way of accurately fitting the SSFR distribution while avoiding overfitting}.
} 
\label{fig:pssfr_gmm_inst}
\end{center}
\end{figure*}

\begin{figure*}
\begin{center}
\includegraphics[width=0.8\textwidth]{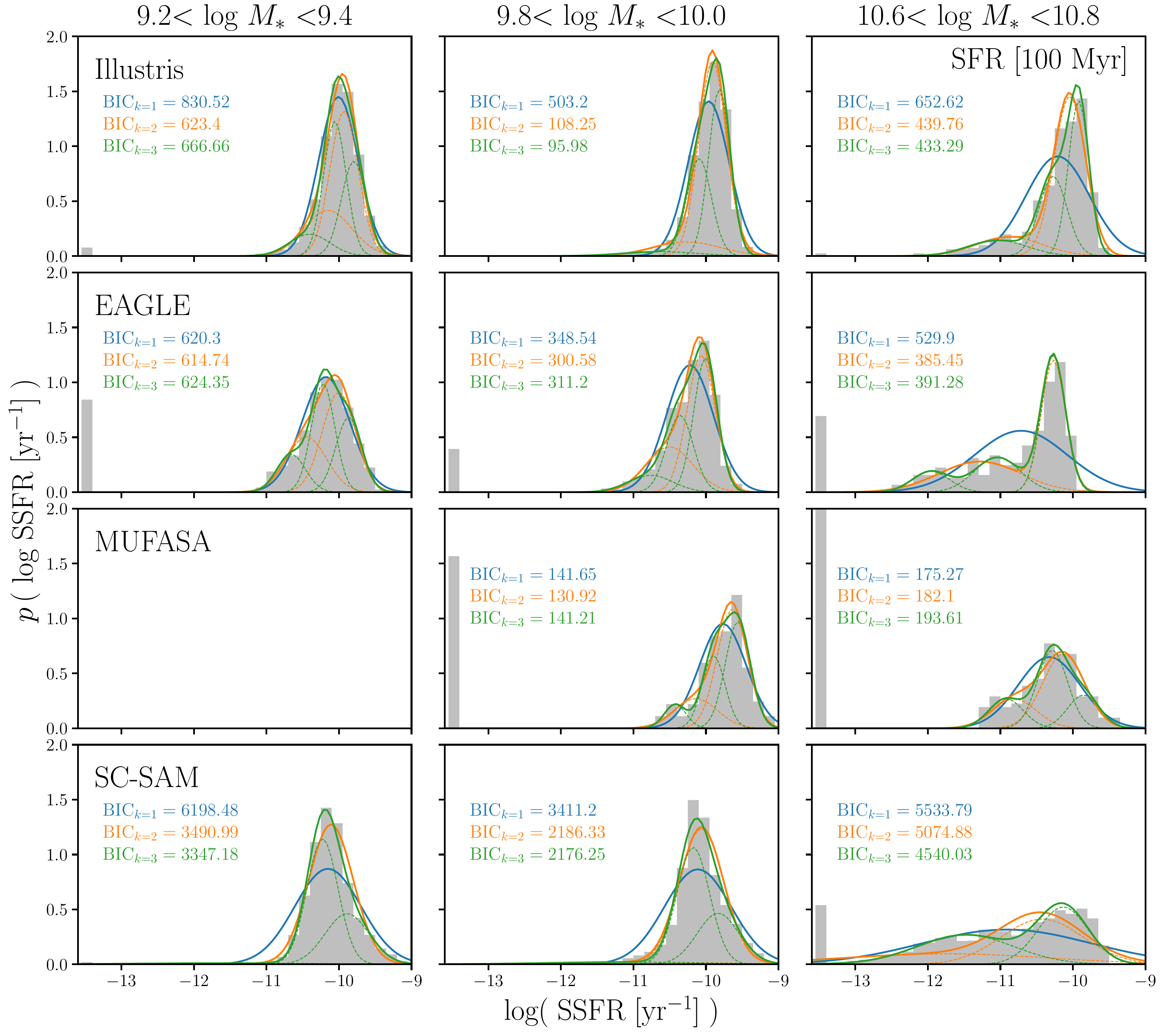} 
\caption{Same as Figure~\ref{fig:pssfr_gmm_inst} but for the $100\,\mathrm{Myr}$
SSFR distributions.} 
\label{fig:pssfr_gmm_100myr}
\end{center}
\end{figure*}

In order to derive the best-fit GMM used for identifying the SFS in 
each $M_*$ bin, we compare GMM fits with $k\leq 3$ components using
their BICs (Section~\ref{sec:sfmsfit}). In Figures~\ref{fig:pssfr_gmm_inst} 
and~\ref{fig:pssfr_gmm_100myr}, we illustrate this comparison among the  
GMMs with $k=1, 2,$ and $3$ (blue, orange, and green) components fit
to the instantaneous SSFR distributions, $P(\log\,\mathrm{SSFR})$, of the 
Illustris, EAGLE, {\sc Mufasa}, and SC-SAM (top to bottom panels) centrals 
in three stellar mass bins: 
$9.2 <\log\,M_*<9.4$, $9.8 <\log\,M_*<10$, and $10.6 <\log\,M_*<10.8$ (left to right). 
The SSFR distributions in Figures~\ref{fig:pssfr_gmm_inst} 
and~\ref{fig:pssfr_gmm_100myr} are derived using instantaneous and 
$100\,\mathrm{Myr}$ SFRs respectively. Galaxies with unmeasurably low SFR 
are represented at the edge of the SSFR distributions with $\log\,\mathrm{SSFR}=-13.5$.
In each panel, we also present the BICs and plot every component of the 
GMM fits (dashed) in their respective colors. These figures illustrate
the advantages of the data-driven GMM based fitting and BIC based
model selection used in our SFS fitting.

For instance, in most of the highest $M_*$ bin (right panels in both figures) 
the $k=1$ GMM fits do not reflect the clearly bimodal SSFR distributions.
In these cases, the $\mathrm{BIC}_{k=1}$ is significantly larger than
$\mathrm{BIC}_{k=2}$ and $\mathrm{BIC}_{k=3}$, so our BIC based model 
selection favors GMMs with more than one component. In fact, GMMs with
more components are more flexible and generally can better fit the underlying 
distribution. However as the EAGLE and {\sc Mufasa} $9.8 <\log\,M_*<10$ 
bins of the figures illustrate, our BIC based model selection does not 
always favor the higher $k$ GMM fits. Although the $k=3$ GMMs have the 
lowest $\chi^2$ in these panels, because of the penalty term for the 
number of model parameters, our BIC criteria favors the $k=2$ GMMs.
According to the BICs, the $k=3$ GMMs in these panels overfit the 
SSFR distributions.


From the best-fit GMMs, we identify the SFS components iteratively starting 
from the lowest $M_*$ bin as described in Section~\ref{sec:gmm}. We 
consider other components, depending on their mean, as intermediate or high 
SF components in Section~\ref{sec:beyondsfms}. The SC-SAM in particular has 
high SF components at $M_* \lesssim 10^{10}M_\sun$ (bottom left and center 
panels of Figures~\ref{fig:pssfr_gmm_inst}  and~\ref{fig:pssfr_gmm_100myr}).
In these cases, the SSFR distribution is not well described by a single 
log-normal distribution. Instead the distribution is asymmetric with a 
heavier tail on the more star-forming end of the distribution. An
extra component to account for the heavier tail improves the fit more 
than the penalty term, giving us the high SF components.



\section{SFR Resolution Effects in Hydrodynamic Simulations} \label{app:zerosfr}
In our analysis, we consistently derive SFRs for all of the simulated
galaxies on two timescales: instantaneous and averaged over 
$100\,\mathrm{Myr}$ (Section~\ref{sec:galsims}). For the hydrodynamic 
simulations, SFR averaged over $100\,\mathrm{Myr}$ is derived using 
the formation times of the star particles in the simulation, which 
means that the mass and temporal resolutions of the simulations 
impact the $100\,\mathrm{Myr}$ SFR. In Illustris, EAGLE, and {\sc Mufasa},
their $100\,\mathrm{Myr}$ SFRs have resolutions of 
$\Delta_\mathrm{SFR} = 0.0126$, $0.018$, and $0.182\,M_\sun \mathrm{yr}^{-1}$, 
corresponding to baryon particle masses of $1.26 \times 10^6\ M_{\sun}$, 
$1.8 \times 10^6\ M_{\sun}$, and $1.82 \times 10^7\ M_{\sun}$, respectively. 
For SFR averaged over $10\,\mathrm{Myr}$, the $\Delta_\mathrm{SFR}$s would be 
10 times larger. Therefore, we instead use instantaneous SFRs to measure 
star formation on the shortest timescale.

\begin{figure*}
\begin{center}
\includegraphics[width=0.7\textwidth]{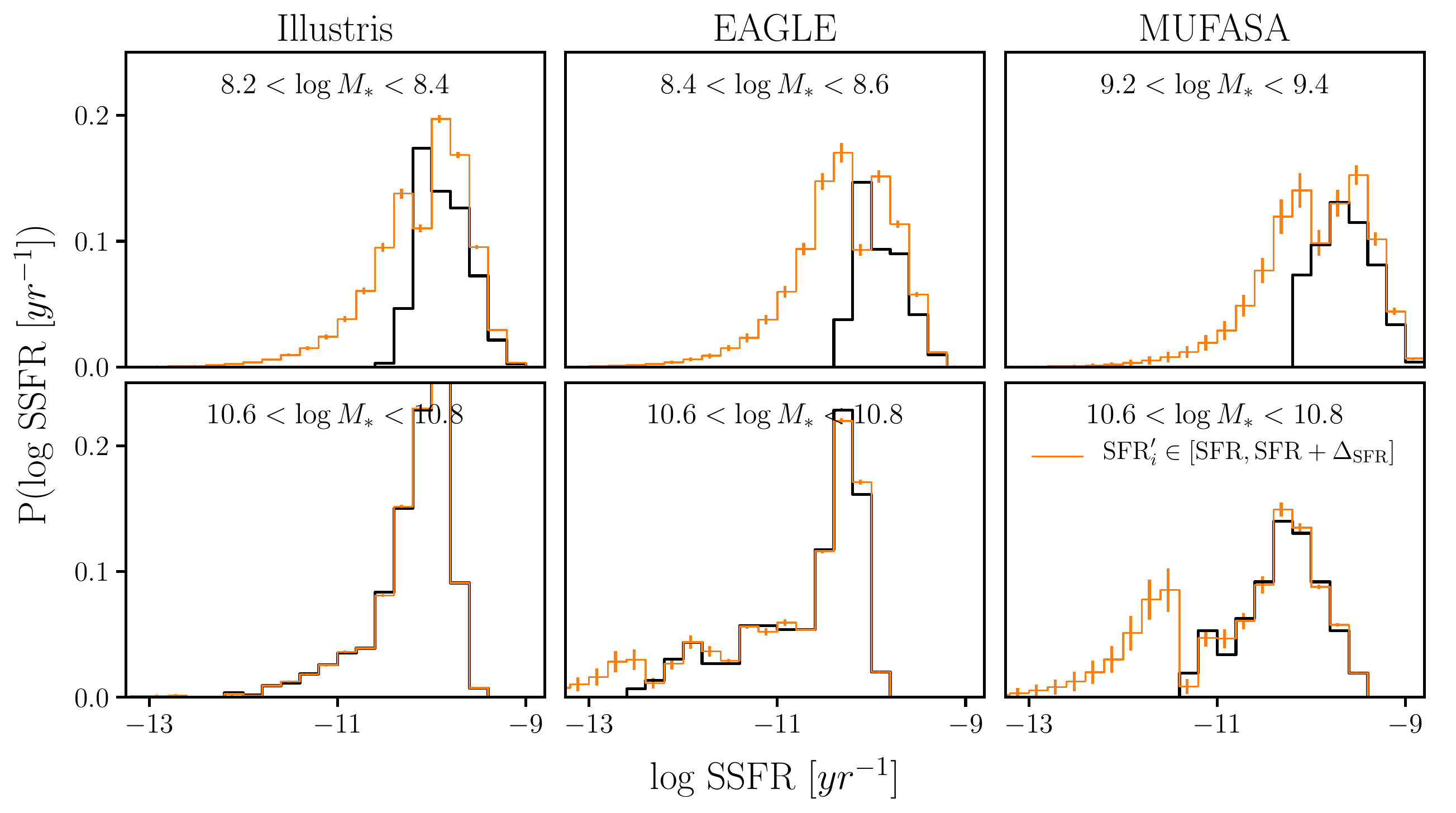} 
\caption{The impact of SFR resolution on the SSFR distribution, 
$P(\log\,\mathrm{SSFR})$, in two stellar mass bins of the hydrodynamic 
simulations:
Illustris (left), EAGLE (center), and {\sc Mufasa} (right). We plot the 
$P(\mathrm{SSFR})$ distributions using the $100\,\mathrm{Myr}$
SFRs \emph{with} resolution effects in black. These exclude galaxies with 
unmeasurably low SFRs. In orange, we plot the $P(\log\,\mathrm{SSFR})$ 
distributions where the SFRs of the galaxies
are sampled uniformly within the SFR resolution range 
($[\mathrm{SFR}_i, \mathrm{SFR}_i+\Delta_\mathrm{SFR}]$). The 
uncertainties for the orange $P(\mathrm{SSFR})$s are estimated from 
re-sampling the SFR of each galaxy based on the SFR resolution. 
At low stellar masses (top) the SFR resolution significantly impacts 
the star-forming end of $P(\mathrm{SSFR})$s. At higher stellar masses, 
although the SFR resolution impacts the $P(\mathrm{SSFR})$s, the effect 
is limited to below $\log\,\mathrm{SSFR} < -11$.
} 
\label{fig:sfrres_pssfr}
\end{center}
\end{figure*}

For galaxies with high $100\,\mathrm{Myr}$ SFR, the resolution 
$\Delta_\mathrm{SFR}$ is relatively small compared to their SFRs and
therefore it does not have a significant impact. However for low SFR 
galaxies, the resolution effect is more significant. At the lowest SFR 
end, galaxies that, without the resolution effect, would have SFR ranging 
$0 < \mathrm{SFR} < \Delta_\mathrm{SFR}$, may have unmeasurably low SFR 
($\mathrm{SFR}{=}0$) with the resolution effect. These galaxies are thereby 
not included in the $\mathrm{SFR}$--$M_*$ plane or when we identify the SFS. In
Figure~\ref{fig:sfrres_pssfr}, we present the impact of excluding these
galaxies and the overall resolution effect on the $P(\log\,\mathrm{SSFR})$ 
distributions of the hydrodynamic simulations in two stellar mass bins. In 
black, we plot the $P(\log\,\mathrm{SSFR})$ distributions using the 
$100\,\mathrm{Myr}$ SFRs \emph{with} resolution effects
(\emph{excluding} galaxies with unmeasurably low SFR). In orange, we plot the 
$P(\log\,\mathrm{SSFR})$ distributions of \emph{all} galaxies where
$\mathrm{SFR}'_i$ of each galaxy sampled uniformly within the SFR resolution range, 
$[\mathrm{SFR}_i, \mathrm{SFR}_i+\Delta_\mathrm{SFR}]$. Uncertainties 
for the orange $P(\log\,\mathrm{SSFR})$s are derived from repeating this 
SFR sampling $100$ times. For the low $M_*$ bins (top), the SFR resolution 
affects the $P(\log\,\mathrm{SSFR})$s well above $\log\,\mathrm{SSFR}{=}{-}11.0$ 
on the star-forming end of the distribution. Meanwhile, the impact at higher
$M_*$ (bottom), is limited to the low SSFR end. 

\begin{figure*}
\begin{center}
\includegraphics[width=0.75\textwidth]{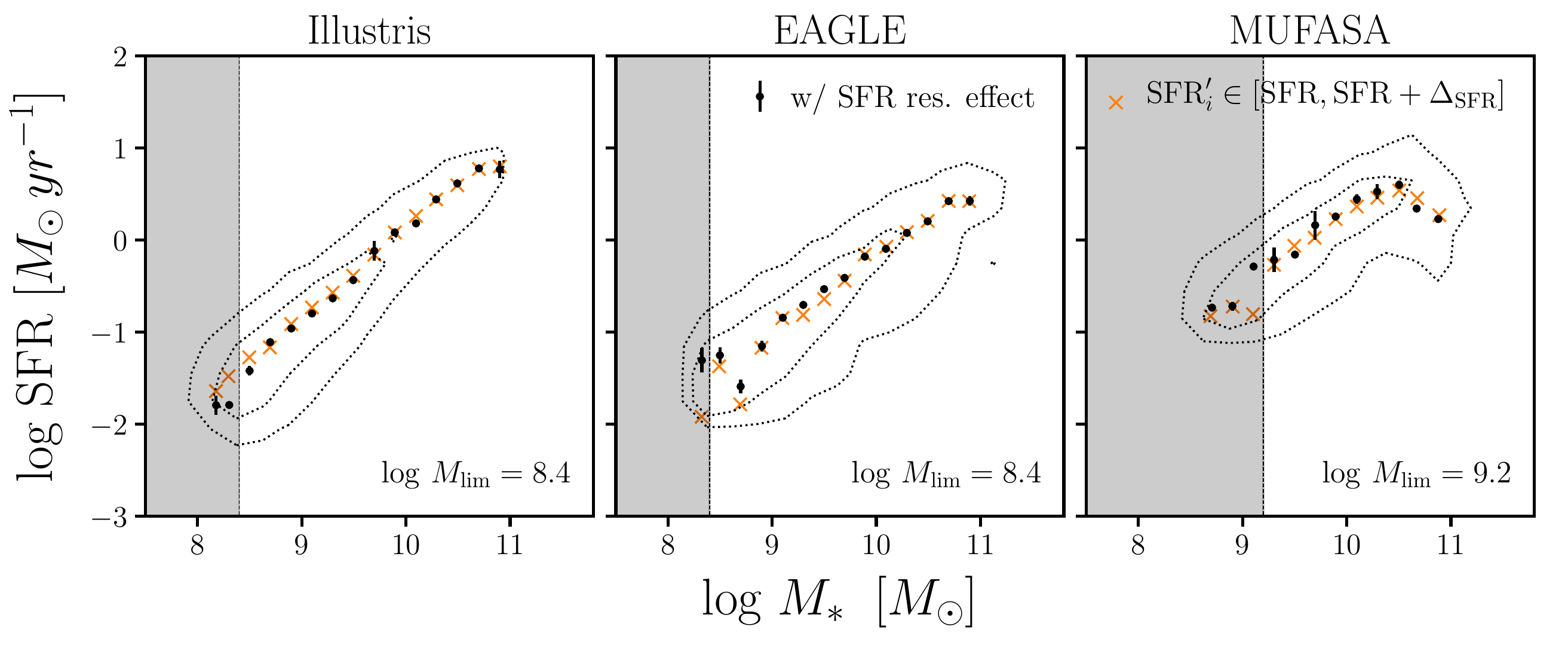} 
\caption{The resolution effect of $100\,\mathrm{Myr}$ SFRs in the hydrodynamic 
simulations (Illustris, EAGLE, and {\sc Mufasa}) impact the identified SFSs 
at low stellar masses. In black we plot the best-fit SFS 
with the resolution effects. In orange we plot the best-fit SFS where 
the SFR for each galaxy is sampled uniformly within the resolution range: 
$\mathrm{SFR}_i^{\prime} \in [\mathrm{SFR}_i, \mathrm{SFR}_i +
\Delta_\mathrm{SFR}]$). Based on the discrepancy between the fits, we determine 
stellar mass limits above which the SFR resolution does {\em not} 
significantly impact ($< 0.2\,\mathrm{dex}$) the identified SFS. For Illustris, 
EAGLE, and {\sc Mufasa} this corresponds to $\log M_\mathrm{lim} = 8.4, 8.4$, 
and $9.2$, as shown in the gray shaded region.} 
\label{fig:mlim_res}
\end{center}
\end{figure*}
In order to better quantify the impact of the SFR resolution effect
on our SFS fitting, in Figure~\ref{fig:mlim_res} we compare the SFS fits 
using $100\,\mathrm{Myr}$ SFRs \emph{with} resolution effects (black) to 
the SFS fits using $100\,\mathrm{Myr}$ SFRs sampled uniformly within the 
SFR resolution range (orange; 
$\mathrm{SFR}_i^{\prime} \in [\mathrm{SFR}_i, \mathrm{SFR}_i + \Delta_\mathrm{SFR}]$). 
The uncertainties of our SFS fits in black are calculated using bootstrap 
resampling (Section~\ref{sec:sfmsfit}). In agreement with 
Figure~\ref{fig:sfrres_pssfr}, we find that the SFR resolution 
significantly impacts the identified SFS at low $M_*$. Moreover, using the 
comparison of Figure~\ref{fig:mlim_res}, we determine the stellar mass 
limit above which the SFR resolution does {\em not} significantly impact 
the identified SFS --- \emph{i.e.} the shift in best-fit SFS is below 
$0.2\,\mathrm{dex}$. For Illustris, EAGLE, and {\sc Mufasa} we determine 
$\log M_\mathrm{lim} = 8.4, 8.4$, and  $9.2$, respectively. For EAGLE, 
where we have a higher resolution box ($8\times$ higher baryon mass 
resolution) available, we further confirm that the SFS is not significantly 
impacted above $\log M_\mathrm{lim}$.

\begin{figure*}
\begin{center}
\includegraphics[width=0.7\textwidth]{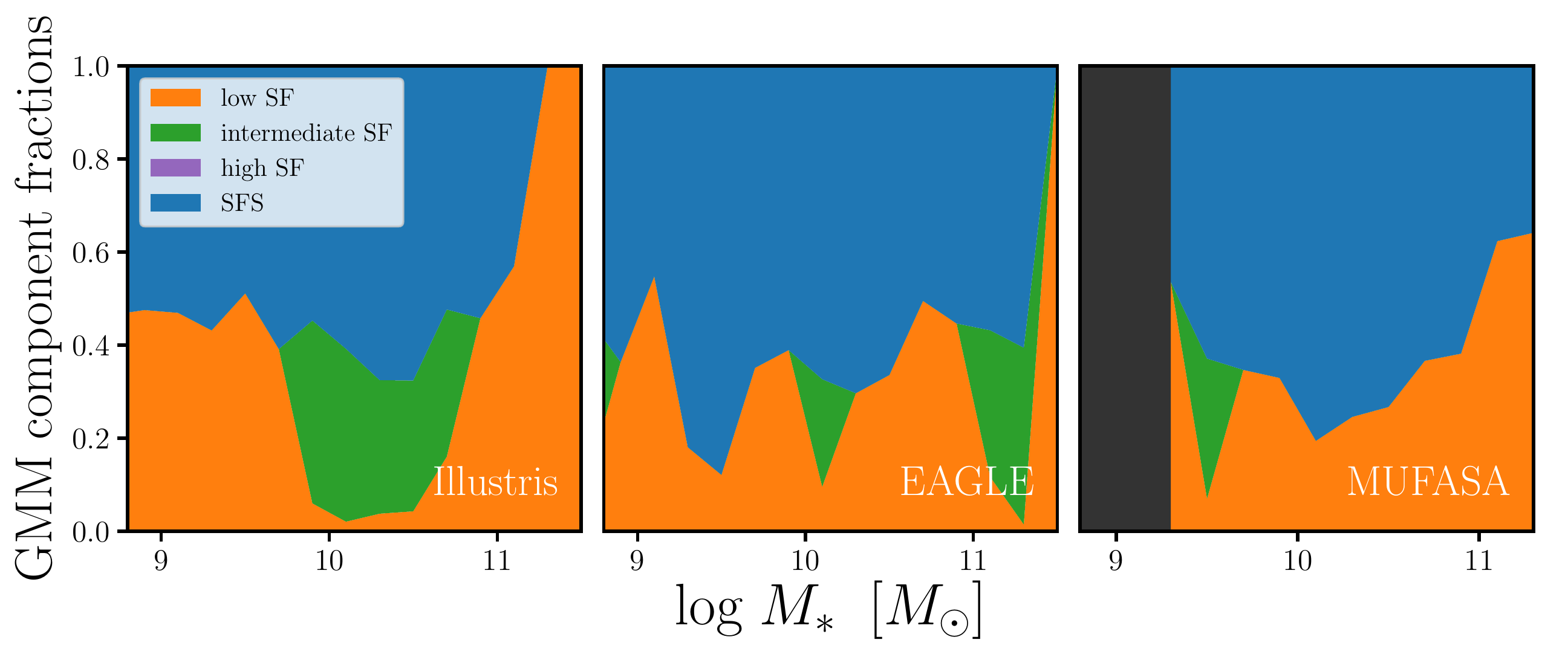}
\caption{Fractional contributions, $\pi_i$, of the best-fit GMM components 
from our method for the hydrodynamic simulations (Illustris, EAGLE, and 
{\sc Mufasa}) where we uniformly sample the $100\,\mathrm{Myr}$ SFRs within
the SFR resolution range --- $\mathrm{SFR}_i^{\prime} \in [\mathrm{SFR}_i, \mathrm{SFR}_i + 
\Delta_\mathrm{SFR}]$. Compared to Figure~\ref{fig:kandinsky}, we find SFR 
resolution has no significant impact on the qualitative results in Section~\ref{sec:beyondsfms}.} 
\label{fig:mlim_fcomp}
\end{center}
\end{figure*}

In addition to its effect on the SFS fits, we also examine the impact of 
SFR resolution on our results regarding the non-SFS components of our GMM 
fitting (Figure~\ref{fig:kandinsky}). In 
Figure~\ref{fig:mlim_fcomp} we present the fraction contributions ($\pi_i$) 
of the best-fit components for the Illustris, EAGLE, and {\sc Mufasa} 
simulations, where we uniformly sample the SFRs within the SFR resolution 
range (same as above). Besides no longer having a component of galaxies 
with unmeasurably low SFRs due to the SFR sampling, we find no significant 
change from the $\pi_i$ of Figure~\ref{fig:kandinsky} and,  thus, the results of Section~\ref{sec:beyondsfms}. 

\bibliographystyle{aasjournal}
\bibliography{paper1}
\end{document}